\documentstyle[12pt,aaspp4,flushrt,amssym]{article}

\begin{document}

\title{The Luminosity Function of Galaxies 
         in the Las Campanas Redshift Survey} 
\author{Huan Lin \altaffilmark{1} and Robert P. Kirshner}
\affil{Harvard-Smithsonian Center for Astrophysics, 60 Garden St.,
       Cambridge, MA 02138, USA \\
       hlin@cfa.harvard.edu, kirshner@cfa.harvard.edu}
\author{Stephen A. Shectman and Stephen D. Landy}
\affil{Carnegie Observatories, 813 Santa Barbara St., Pasadena, CA
       91101, USA \\
       shec@ociw.edu, landy@ociw.edu}
\author{Augustus Oemler}
\affil{Dept. of Astronomy, Yale University, 
       New Haven, CT 06520-8101, USA \\
       oemler@astro.yale.edu}
\author{Douglas L. Tucker}
\affil{Astrophysikalisches Institut Potsdam, An der Sternwarte 16,
       D-14482 Potsdam, Germany \\
       dtucker@aip.de}
\and
\author{Paul L. Schechter}
\affil{Dept. of Physics, Massachusetts Institute of Technology,
       Cambridge, MA 02139, USA \\
       schech@achernar.mit.edu}

\vspace{2cm}

\affil{To appear in {\em The Astrophysical Journal},
             {\bf 464}, June 10, 1996}

\altaffiltext{1}{Present Affiliation: Dept. of Astronomy, University of
    Toronto, 60 St. George St., Toronto, ON M5S 3H8, Canada, 
    lin@astro.utoronto.ca}

\clearpage

\begin{abstract}

We present the $R$-band luminosity function for a sample of 18678 
galaxies, with
average redshift $z = 0.1$, from the Las Campanas Redshift Survey (LCRS).
The luminosity function may be fit by a Schechter function with 
$M^* = -20.29 \pm 0.02 + 5 \log h$, $\alpha = -0.70 \pm 0.05$, and 
$\phi^* = 0.019 \pm 0.001 \ h^3$~Mpc$^{-3}$, for absolute magnitudes 
$-23.0 \leq M - 5 \log h \leq -17.5$ and 
$h = H_0 / (100$~km~s$^{-1}$~Mpc$^{-1})$. 
Over the same absolute magnitude range, the mean galaxy density is
$0.029 \pm 0.002 \ h^3$~Mpc$^{-3}$ for a volume extending to 
$cz = 60000$~km~s$^{-1}$.

We compare our luminosity function to that from other redshift surveys;
in particular our luminosity function normalization is consistent with
that of the Stromlo-APM survey, and is therefore a factor of two below 
the normalization implied by the $b_J \approx 20$ bright galaxy counts.
Our normalization thus indicates that much more evolution is needed to match the
faint galaxy count data, compared to minimal evolution models which 
normalize at $b_J \approx 20$.
Also, we show that our faint-end
slope $\alpha = -0.7$, though ``shallower'' than typical previous values
$\alpha = -1$, results primarily from fitting the detailed shape of the 
LCRS luminosity function, rather than from any absence of intrinsically
faint galaxies from our survey. 

Finally, we find that the faint end of the 
luminosity function is dominated by galaxies with emission lines. 
Using [OII] 3727 equivalent width $W_{\lambda} = 5$~\AA \ as the dividing line, 
we find significant 
differences in the luminosity functions of emission and non-emission galaxies, 
particularly in their $\alpha$ values; emission galaxies have Schechter
parameters $M^* = -20.03 \pm 0.03 + 5 \log h$ and $\alpha = -0.9 \pm 0.1$, 
while non-emission galaxies are described by 
$M^* = -20.22 \pm 0.02 + 5 \log h$ and 
$\alpha = -0.3 \pm 0.1$.
The average [OII] 3727 equivalent widths do not change significantly 
with redshift, consistent with a star formation
rate which stays constant over the depths sampled by the LCRS.
This result holds for galaxies of different luminosities, and 
over the respective redshift ranges that these
galaxies may be observed, in particular up to about $z = 0.2$ for galaxies
brighter than $M^*$. 

\end{abstract}


\clearpage

\section{Introduction}

The galaxy luminosity function is one of the most basic and fundamentally
important quantities in observational cosmology. For example, it is 
essential for the analysis of large-scale structure statistics from
redshift surveys, and knowledge of the local luminosity function is
necessary for the proper interpretation of the faint galaxy count
data and galaxy evolution at high redshift (e.g., \cite{koo92}). 

Though much work has gone into determination of the local $z = 0$ optical
luminosity function (see \cite{fel77}; \cite{2:KOSII}, 1983; \cite{eep};
\cite{bin88}; and recent work by \cite{lov92}; \cite{marz94}), there is
still some uncertainty in its faint end as well as in its
normalization (e.g., \cite{coll95}). In this paper we investigate
the local luminosity function by presenting results for
a sample of 18678 Las Campanas Redshift Survey (LCRS) galaxies,
selected from CCD photometry in a ``hybrid'' red Kron-Cousins $R$-band
(see \S~\ref{lumdata} below).
This is the largest galaxy sample for which the luminosity function and mean 
density have been computed. The combination of the mean survey depth 
of $z = 0.1$ and the large survey volume 
($1.0 \times 10^7 \ h^{-3}$~Mpc$^3$ out to $cz = 60000$ km~s$^{-1}$)
imply that our results are not biased by local inhomogeneities. 
In addition, the spectra of LCRS galaxies
allow us to study the luminosity function for samples divided by
[OII] 3727 emission strength, a rough indicator of the star-formation rate
(\cite{ken92}).

We describe relevant aspects of the LCRS construction in \S~\ref{lumdata}. 
We then 
detail our methods for measuring the galaxy luminosity function and mean 
density in \S~\ref{methods}. Our results are presented in \S~\ref{lfrholcrs}, 
and we compare them to those of other redshift surveys in \S~\ref{comparisons}. 
Some implications of
our results for faint galaxy counts and galaxy evolution are discussed in
\S~\ref{counts}. We summarize our conclusions in \S~\ref{conclusions}.

\section{The Las Campanas Survey Data} \label{lumdata}

We begin by describing the Las Campanas survey parameters; 
detailed descriptions of the LCRS may be found in Shectman et al. 
(1992, 1995, 1996), Tucker (1994), and Lin (1995). 
There are 23690 galaxies, with an average 
redshift $z = 0.1$, which lie within the survey photometric limits 
and geometric borders considered in this paper. 
The survey galaxies were selected from a 
CCD photometric catalog obtained from drift scans taken on the
Las Campanas 1.0~m Swope telescope with a Gunn $r$ filter (\cite{thu76}).
Follow-up spectroscopy was obtained using the 
multi-object fiber-optic spectrograph on the Las Campanas 2.5~m DuPont 
telescope. The survey geometry is that of six ``slices,'' $1.5\arcdeg$ in 
declination by $80\arcdeg$ in right ascension, three each in the North and South
galactic caps.

The survey data may be divided into two main parts: the first 20\% of the data
was obtained using a 50-object fiber system, while the remaining 80\% of the
data was taken with a 112-object system. The survey slices were built up
by observing $1.5\arcdeg$ by $1.5\arcdeg$ spectroscopic fields, with
a maximum of 50 or 112 galaxies observed per field. The survey photometric
limits were chosen so that there would be typically more targets 
per field than available fibers, and we selected targets at random among
those that met the selection criteria.
Because we generally do not re-observe any of 
our fields, we keep track of the variable field-to-field sampling 
fractions $f_i$ in our subsequent statistical analyses.

To be precise, though our photometry was obtained through a Gunn $r$
filter, the calibration was done relative to standard stars
(\cite{gra81}; \cite{gra82}) in the
Kron-Cousins $R$-band. Our magnitudes are thus in a ``hybrid'' Kron-Cousins
$R$ system. Nevertheless, the difference between our hybrid 
Kron-Cousins $R$ system and the true one is small; 
comparison of photometry of LCRS galaxies obtained in both systems
yields a zero-point difference of $< 0.1$
mag and a color term of $< 0.2$ mag per magnitude of the
color (Johnson $V -$ Kron-Cousins $R$). See Tucker (1994) and 
Shectman et al.\ (1996) for more details.

\begin{figure}
\plotfiddle{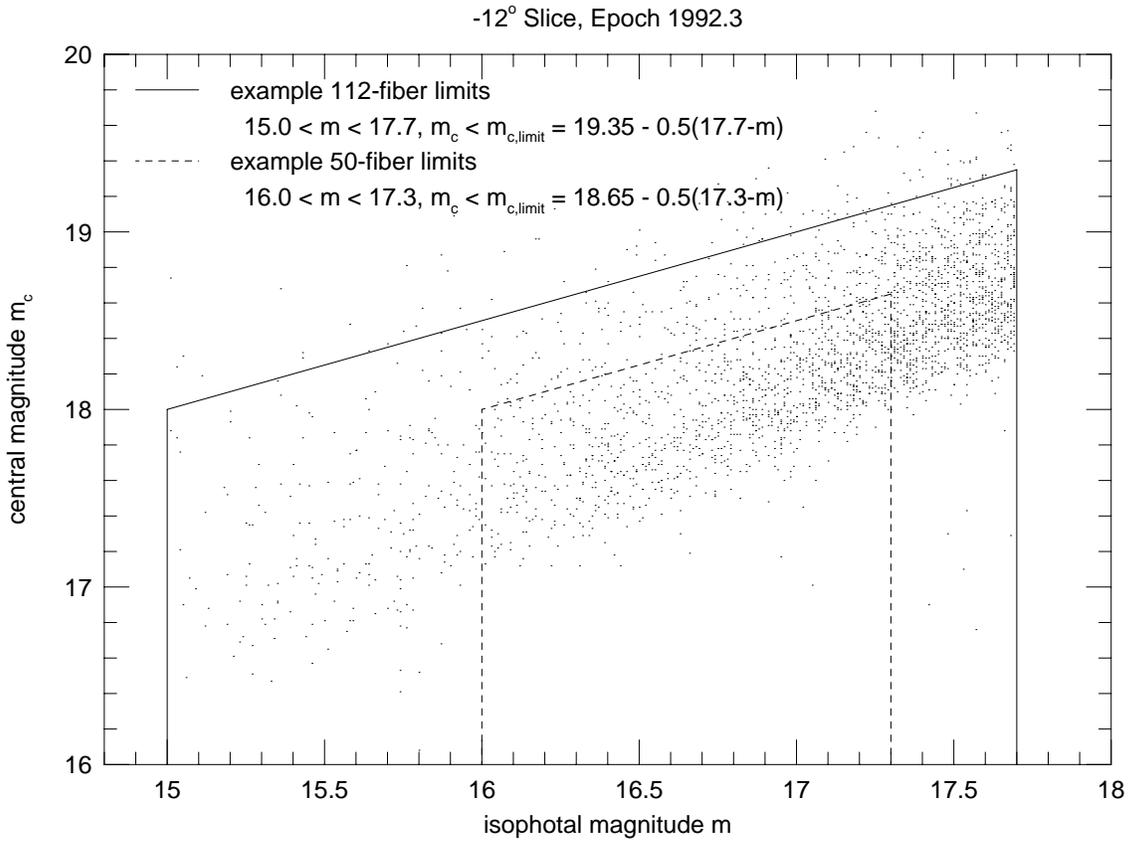}{13cm}{0}{62}{62}{-250}{0}
\caption{Examples of the photometric selection criteria in the
         isophotal magnitude ($m$) - central magnitude ($m_c$) plane
         applied to the 112- and 50-fiber LCRS data.}
\label{figsel}
\end{figure}

Objects were selected for the survey using isophotal magnitudes $m$, applying
both faint and bright isophotal magnitude limits. The limiting isophote
is 15\% of sky, or approximately 23 mag arcsec$^{-2}$.
In addition, a central magnitude $m_c$, approximately
measuring flux that would enter through the $3.5\arcsec$-diameter aperture of
a fiber, is also computed for each object. An additional cut line in the
$m$-$m_c$ plane is then imposed in order to exclude low central surface 
brightness objects, for which redshifts are more difficult to obtain.
For the 50-fiber data, the nominal isophotal magnitude
limits are $16.0 \leq m < 17.3$, and the central surface brightness cut 
excludes about 20\% of objects otherwise meeting the isophotal limits.
For the 112-fiber data, the nominal isophotal limits are $15.0 \leq m < 17.7$,
with exclusion of just the lowest 4-9\% of objects by surface brightness.
We illustrate our isophotal and central magnitude selection criteria for
a sample of LCRS galaxies in Figure~\ref{figsel}.
There are small field-to-field variations from the nominal photometric limits;
these are accounted for in our analyses. We will find it convenient
to divide our data into four subsets by galactic cap, North and South, 
and by fiber number, 50 and 112 fibers; we will refer to these subsets as
N50, S50, N112, and S112. Table~\ref{tabsamps} summarizes details concerning
these four samples. Also, it turn outs that our fiducial sample will be
the combined North plus South 112-fiber data set, which we will refer
to as NS112.

Throughout this paper we adopt a Hubble constant 
$H_0 = 100 \ h \ {\rm km \ s}^{-1}$~Mpc$^{-1}$ and a deceleration parameter 
$q_0 = 0.5$. Where the $h$ dependence is not explicitly
indicated, we have used $h = 1$.
We also assume a uniform Hubble flow so that the absolute 
magnitude $M$ of a galaxy of isophotal magnitude $m$ and redshift $z$ is
given by
\begin{equation}
M = m - 25 - 5 \log d_L - k(z) \ ,
\end{equation}
where the luminosity distance $d_L$ is
\begin{equation}
d_L = \frac{c}{H_0 q_0^2} 
      \left[ q_0 z + (1-q_0) (1-\sqrt{1+2 q_0 z}) \right] \ ,
\end{equation}
and where the $k$-correction $k(z)$ is taken to be
\begin{equation}
k(z) = 2.5 \log(1+z) \ ,
\end{equation}
as appropriate in the mean for an $r$-band-selected galaxy sample
(\cite{tuck94}; \cite{2:KOSSII}).
Note that galaxy peculiar velocities, of order 
$\sigma_{12} / \sqrt{2} \approx 350$ km~s$^{-1}$ ($\sigma_{12}$ is the
{\em pairwise} peculiar velocity dispersion; see \cite{lin96sig}; \cite{marz95}),
and our velocity measurement errors, on average 67 km~s$^{-1}$ 
(\cite{lin95the}), are both quite small compared to the mean survey redshift 
$cz = 30000$ km~s$^{-1}$. We therefore neglect the effect of peculiar
velocities and velocity errors in the calculation of absolute
magnitudes. Similarly, the effect of using a different value of $q_0$ is also
small: $M({q_0}') - M(q_0=0.5) \approx ({q_0}'-0.5) z$, e.g.,
$-0.04$ magnitude at our average $z = 0.1$ for the case ${q_0}' = 0.1$.

\section{Methods} \label{methods}

\subsection{Luminosity Function} \label{methods:LF}

We compute the luminosity function using two related methods which are
unbiased by density inhomogeneities in the galaxy distribution. These are
the parametric 
maximum-likelihood method of Sandage, Tammann, \& Yahil (1979; 
hereafter STY) and the non-parametric stepwise maximum-likelihood
(SWML) method of Efstathiou, Ellis, \& Peterson (1988; hereafter
EEP). We describe these methods below, and we will discuss some modifications 
needed in the analysis of the LCRS sample in \S~\ref{methods:modifications}.

Consider a galaxy $i$ observed at redshift $z_i$ in a flux-limited redshift
survey like the LCRS. Let $m_{{\rm min},i}$ and $m_{{\rm max},i}$ denote 
the apparent
magnitude limits of the field in which galaxy $i$ is located. Let $\phi(M)$
be the differential galaxy luminosity function which we want to determine. 
Then the probability that 
galaxy $i$ has absolute magnitude $M_i$ is given by
\begin{equation} \label{eqprob}
p_i \equiv p(M_i \vert z_i) = 
    \phi(M_i) \left/ \int^{M_{\rm max}(z_i)}_{M_{\rm min}(z_i)}
                             \phi(M) dM \right. \ ,
\end{equation}
where the absolute magnitude limits at $z_i$ are
\begin{equation}
{ \left\{ \begin{array}{c}
           M_{\rm min}(z_i) \\
           M_{\rm max}(z_i)
          \end{array} \right\} } = 
{ \left\{ \begin{array}{c}
           m_{{\rm min},i} \\
           m_{{\rm max},i}
          \end{array} \right\} } 
- 25 - 5 \log {d_L}_i - k(z_i) \ .
\end{equation}
The luminosity distance $d_L$ and the $k$ correction $k(z)$ for 
LCRS galaxies are computed as described in \S~\ref{lumdata}.
We can then form a likelihood function $\frak L$ for having a survey of $N$ galaxies,
with respective absolute magnitudes $M_i$, by multiplying the probabilities
$p_i$:
\begin{equation} 
\frak L = p(M_1, \ldots , M_N \vert z_1, \ldots , z_N)
  = \prod_{i=1}^{N} p_i \ ,
\end{equation}
or
\begin{equation} \label{eqlnlike}
\ln \frak L = \sum_{i=1}^{N} \left\{ 
                         \ln \phi(M_i) - 
                         \ln \int^{M_{\rm max}(z_i)}_{M_{\rm min}(z_i)} 
                           \phi(M) dM
                       \right\} \ .
\end{equation}
In the STY method one assumes a parametric model for $\phi(M)$, and the
parameters describing $\phi(M)$ are determined by maximizing the likelihood 
$\frak L$, or equivalently $\ln \frak L$,
with respect to those parameters. In our case we take as our model for
$\phi(M)$ the Schechter function (\cite{sch76})
\begin{equation} \label{eqschech}
\phi(M) = (0.4 \ln 10) \ \phi^* \ [10^{0.4(M^*-M)}]^{1+\alpha} \
                         \exp [-10^{0.4(M^*-M)}] \ ,
\end{equation}
and use the STY method to find the characteristic magnitude $M^*$ and the 
faint-end slope $\alpha$.
The normalization $\phi^*$ drops out in 
equation~(\ref{eqprob}) and has to be determined using a different method to be 
described below in \S~\ref{methods:dens}. 
Also, error ellipses in the $M^*$-$\alpha$ plane 
may be drawn by finding the contour corresponding to
\begin{equation} \label{ellipse}
\ln \frak L = \ln \frak L_{\rm max} - \case{1}{2} \Delta \chi^2 \ ,
\end{equation}
where $\Delta \chi^2$ is the change in $\chi^2$ appropriate for the 
desired confidence level
and a $\chi^2$ distribution with 2 degrees of freedom.

Alternatively, one does not have to assume a particular functional form
for $\phi(M)$. Rather, in the EEP stepwise maximum likelihood method (SWML), 
the luminosity function is taken to be a series of $N_p$ steps, each of width
$\Delta M$ in absolute magnitude:
\begin{equation}
\phi(M) = \phi_k \ , \ \ M_k - \Delta M / 2 < M < M_k + \Delta M / 2 \ , \ \
                     k = 1, \ \ldots , \ N_p \ .
\end{equation}
In this case equation~(\ref{eqlnlike}) may be rewritten as
\begin{eqnarray}
\ln \frak L & = & \sum_{i=1}^{N} \sum_{k=1}^{N_p} W(M_i - M_k) \ \ln \phi_k 
          \nonumber \\
      & - & \sum_{i=1}^{N} \ln 
            \left\{ \sum_{k=1}^{N_p} \phi_k \ \Delta M \
                    H [M_k, M_{\rm min}(z_i), M_{\rm max}(z_i)]
            \right\} \ , \label{eqsteplnL}
\end{eqnarray}
where the functions $W$ and $H$ are given by
\begin{equation}
W(M_i - M_k) = \left\{ \begin{array}{ll}
                        1, & M_k - \Delta M / 2 \leq M_i 
                                                \leq M_k + \Delta M / 2 \\
                        0, & {\rm otherwise} 
                       \end{array}
               \right. \ ,
\end{equation}
\begin{eqnarray}
\lefteqn{ H [M_k, M_{\rm min}(z_i), M_{\rm max}(z_i)] = } \nonumber \\ 
& & \left\{ \begin{array}{l}
              \min [M_k + \Delta M / 2, M_{\rm max}(z_i)]
            - \max [M_k - \Delta M / 2, M_{\rm min}(z_i)] \ , \\
              \ \ \ {\rm if} \ M_k + \Delta M / 2 \geq M_{\rm min}(z_i) 
                       \ {\rm and} \
                       M_k - \Delta M / 2 \leq M_{\rm max}(z_i) \\
              0, \ \ {\rm otherwise}
            \end{array}
    \right. \ .
\end{eqnarray}
We then maximize $\ln \frak L$, equation~(\ref{eqsteplnL}), with respect to
the parameters $\phi_k$, and subsequently solve for them iteratively using
\begin{eqnarray}
\phi_k \ \Delta M & = & \left. \left\{ 
                         \sum_{i=1}^{N} W(M_i - M_k) 
                        \right\} \ \right/ \\
    & & \sum_{i=1}^{N} 
         \left\{ H[M_k, M_{\rm min}(z_i), M_{\rm max}(z_i)] 
                 \left/ \sum_{j=1}^{N_p} \phi_j  \Delta M \
                        H [M_j, M_{\rm min}(z_i), M_{\rm max}(z_i)]
                 \right. 
         \right\} \nonumber
\end{eqnarray}
As in the STY method, the normalization of the $\phi_k$ is not determined, so
we introduce a constraint similar to that given by EEP to fix the 
normalization:
\begin{equation}
g \equiv \sum_{k=1}^{N_p} \phi_k \ \Delta M \ V(M_k) / V(M_f) - 1 = 0 \ ,
\end{equation}
where $V(M)$ gives the volume (including relativistic effects) over which a 
galaxy of absolute magnitude $M$ can be seen in a flux-limited survey, and 
$M_f$ is a fiducial absolute magnitude, which we take to be 
$-20 \approx M^*$ as appropriate for the LCRS. Also following EEP, we can
estimate the variances of the parameters $\phi_k$ from the first $N_p$
diagonal elements of the covariance matrix given by
\begin{equation}
{\rm cov}(\phi_k) = {\bf \rm I}^{-1}(\phi_k) \ ,
\end{equation}
where ${\bf \rm I}(\phi_k)$ is the information matrix 
\begin{equation}
{\bf \rm I}(\phi_k) =
  - \left. \left[
           \begin{array}{cc}
             \partial^2 \ln \frak L / \partial\phi_i \partial\phi_j +
             (\partial g / \partial \phi_i)
             (\partial g / \partial \phi_j) & \partial g / \partial \phi_j \\
             \partial g / \partial \phi_i & 0 
           \end{array} 
  \right] \right\vert_{\phi_i, \phi_j = \phi_k} \ .
\end{equation}

We can test the goodness of fit of the STY solution by comparing it to the
SWML solution using a likelihood ratio test described by EEP. Let $\frak L_2$ be
the likelihood for the SWML solution given by equation~(\ref{eqsteplnL}), 
and let $\frak L_1$ be the likelihood, again from equation~(\ref{eqsteplnL}), but
with $\phi_k$ set to $\phi_{{\rm STY},k}$, where
\begin{equation}
\phi_{{\rm STY},k} = \int^{M_k+\Delta M/2}_{M_k-\Delta M/2}
                      \phi_{\rm STY}^2(M) V(M) dM
                   \left/
                     \int^{M_k+\Delta M/2}_{M_k-\Delta M/2}
                      \phi_{\rm STY}(M) V(M) dM
                   \right. \ .
\end{equation}
As discussed in EEP, here we use this definition of $\frak L_1$ for the STY solution,
rather than equation~(\ref{eqlnlike}), in order to match the binning present 
in equation~(\ref{eqsteplnL}) but absent from equation~(\ref{eqlnlike}).
EEP show that the likelihood ratio $2 \ln(\frak L_1 / \frak L_2)$ 
approximately follows a
$\chi^2$ distribution with $N_p-3$ degrees of freedom, giving us a means for
testing the quality of the STY solution. 

Alternatively, we can also check the goodness of the STY fit by comparing 
the observed absolute magnitude distribution $d N(M)$ against the distribution
$d N_{\rm STY}(M)$ expected from the STY solution and the observed redshift 
distribution of the $N$ survey galaxies (\cite{yah91}):
\begin{equation}
d N_{\rm STY}(M) = \sum_{i=1}^{N} p(M \vert z_i) dM =
  \sum_{i=1}^{N} \left\{
    \phi(M) dM \left/ \int^{M_{\rm max}(z_i)}_{M_{\rm min}(z_i)}
                              \phi(M) dM \right.
                                 \right\}  \ .
\end{equation}
A $\chi^2$ test can then be used to assess how well the STY prediction matches
the observations.

\subsection{Modifications} \label{methods:modifications}

For thoroughness in the analysis of the LCRS data, we need to consider 
four additional details which will modify the basic methods described
above to calculate the luminosity function.
The most important effect is that of central
surface brightness selection, while the others make fairly small changes in 
$\phi(M)$.

{\it (1) Magnitude Errors.}
From repeat observations we find that our rms isophotal magnitude error
is $\sigma = 0.1$ magnitude. We make the rough approximation
that the magnitude errors are Gaussian distributed with dispersion
$\sigma$, so that
the observed luminosity function $\phi_{obs}$ is a convolution of the 
true luminosity function $\phi$ with a Gaussian magnitude error 
distribution (EEP). Thus
\begin{equation} \label{eqschechconv}
\phi_{obs} = \frac{1}{\sqrt{2 \pi} \sigma}
    \int^\infty_{-\infty} \phi(M') e^{-(M'-M)^2/2\sigma^2} dM' \ ,
\end{equation}
and we use $\phi_{obs}$ in place of $\phi$ in the parametric STY equations.
For simplicity, we will not make this substitution in the non-parametric
SWML equations. Accounting for this dispersion of 0.1 magnitude makes
small systematic 
changes in the best-fit values of the Schechter parameters for our sample: 
$\Delta M^* \approx +0.03$ and $\Delta \alpha \approx +0.03$.

{\it (2) Variable Field Sampling Fractions.}
To account for this effect we weight galaxy $i$'s entry in the likelihood
equations~(\ref{eqlnlike}) and (\ref{eqsteplnL}) by the inverse of the sampling
fraction $f_i$ of the field containing that galaxy (\cite{zuc94}):
\begin{equation}
\ln p_i \longrightarrow \frac{\ln p_i}{f_i} \ .
\end{equation}
Including this effect changes the Schechter parameters very little: both
$\vert \Delta M^* \vert$ and $\vert \Delta \alpha \vert$ are $< 0.01$.

\begin{figure}
\plotfiddle{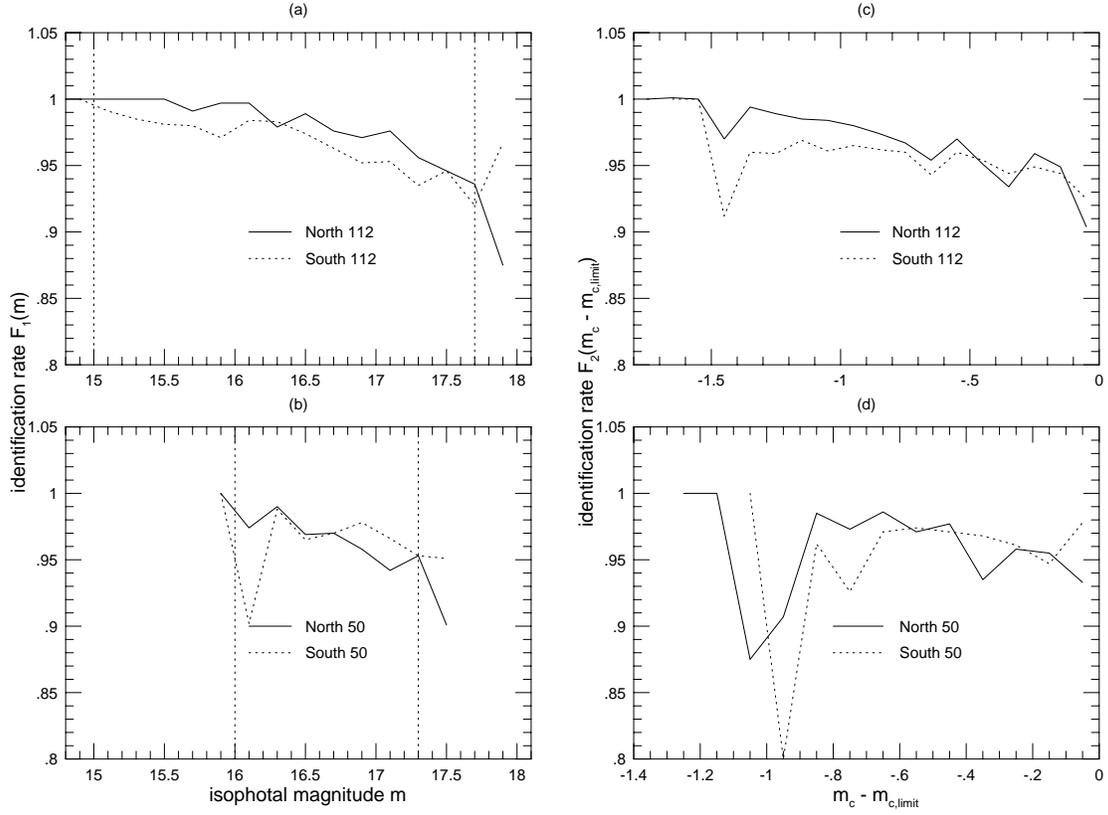}{13cm}{0}{62}{62}{-250}{0}
\caption{(a-b) The identification rate (galaxies plus stars) $F_1(m)$ as a 
          function 
          of apparent magnitude $m$ for the four main subsets of LCRS data.
          Vertical lines show the nominal isophotal magnitude limits.
          (c-d) The identification rate 
          $F_2(m_c-m_{c,limit})$ as a function of ``distance'' 
          $m_c-m_{c,limit}$ from the central magnitude limit.}
\label{figF12}
\end{figure}

{\it (3) Apparent Magnitude and Surface Brightness Incompletenesses.}
Our success rate in identifying spectra as galaxies and stars drops
slightly at fainter apparent magnitudes and also drops slightly close to 
the limiting central surface brightness cut line (Figure~\ref{figsel}). 
We define a completeness
factor $F(m, m_c-m_{c,limit})$, which is a function of isophotal 
magnitude $m$ and of ``distance'' $m_c-m_{c,limit}(m)$ from the central 
magnitude cut line. We take $F$ to be the fraction of spectra we identified 
as galaxies
or stars in each bin on a grid of $m$ and $m_c-m_{c,limit}(m)$ values.
In Figure~\ref{figF12}, we plot the two orthogonal projections
$F_1(m)$ and $F_2(m_c-m_{c,limit})$ of the full
completeness function $F$ for each of the N50, S50, N112, and S112 samples.
From the figure we note that the completeness does not vary much with 
either apparent magnitude or surface brightness; the completeness only drops
to about 0.95 at the faint isophotal magnitude limits or at the central surface
brightness limits. To account for this effect, we weight galaxy $i$'s entry
in the STY and SWML likelihood equations by the inverse of the value of $F$ 
appropriate for that galaxy's isophotal and central magnitudes:
\begin{equation}
\ln p_i \longrightarrow \frac{\ln p_i}{F(m_i, {m_c}_i-{m_{c,limit}}_i)} \ .
\end{equation}
The change introduced is negligible for 
$M^*$, $\vert \Delta M^* \vert < 0.01$, but $\alpha$ does change systematically
by a small amount, $\Delta \alpha \approx -0.04$. (The greater
incompleteness at fainter $m$ translates to a bias against intrinsically
fainter galaxies and hence a more positive $\alpha$, if left uncorrected.)

{\it (4) Central Surface Brightness Selection.}
We empirically find in the LCRS a correlation between a galaxy's absolute
luminosity and that galaxy's distance from our central surface brightness
cut line. We demonstrate this by culling out galaxies from our N112
subset to match the more restrictive surface brightness cuts of the 
N50/S50 and S112 subsets (\S~\ref{lumdata} and Table~\ref{tabsamps}). 
Figure~\ref{figG(M)} shows that the effect of the 
more stringent surface brightness cuts is to reduce the sampling 
completeness $G(M)$ as the absolute magnitude $M$ becomes fainter, where
$G(M)$ is simply the fraction of galaxies at $M$ which remain after the 
low-surface-brightness galaxies have been removed. For the N50, S50, and 
S112 samples, we can
account for this bias against faint galaxies by weighting galaxy $i$'s
entry in the likelihood equations by the inverse of $G$ at
that galaxy's absolute magnitude $M_i$:
\begin{equation}
\ln p_i \longrightarrow \frac{\ln p_i}{G(M_i)} \ .
\end{equation}
We thus correct the surface brightness sampling of the
N50, S50, and S112 subsets to that of the N112 subset.
For the 50-fiber data samples, $\Delta M^* \approx -0.04$ and 
$\Delta \alpha \approx -0.15$; for the S112 sample, $\Delta M^* = -0.01$
and $\Delta \alpha = -0.06$. We do not attempt to
further extrapolate and correct for the N112 sample's own central surface
brightness cut, though we expect the change to be 
$\vert \Delta \alpha \vert < 0.1$ from the example of the S112 data set.

We take account of the latter three effects (2-4) together by the replacement
\begin{equation}
\ln p_i \longrightarrow W_i \ln p_i , \
    W_i \equiv \frac{1}{f_i \ F(m_i, {m_c}_i-{m_{c,limit}}_i) \ G(M_i)} \ ,
\end{equation}
that is, we weight by the inverse of the product of the three completeness
factors. However, because the weighting factors $W_i$ are greater than one,
the likelihood $\ln \frak L$ will be larger in absolute value than when no 
weightings are applied. The likelihood will thus resemble that of a 
larger sample, leading to systematic underestimates in the confidence
ellipses and errors 
we compute (\cite{gla95c}). To remedy this, we renormalize the weights
$W_i$ so that $\sum_{i=1}^N W_i = N$, where $N$ is the number of galaxies
in the sample of interest. The original weights are multiplied by 
renormalization factors of 0.4 to 0.7 depending on the data set, 
in particular by 0.68 in the case of our fiducial NS112 sample.
We have verified that this procedure yields 
error estimates similar to those for the case when no weights are applied.

\subsection{Mean Density} \label{methods:dens}

To calculate the mean galaxy density $\bar\rho$, one can use the 
minimum-variance estimator of Davis and Huchra (1982) (also \cite{lov92}).
We can adapt it to the LCRS as follows. For galaxies $i$ within redshift
limits $z_1 < z_i < z_2$ and absolute magnitude limits $M_1 < M_i < M_2$, we
have
\begin{equation} \label{eqrho}
\bar\rho = \frac{ \sum_i w(z_i) \ W_i }
                { \int^{z_2}_{z_1} S(z) w(z) (\frac{dV}{dz}) dz } \ ,
\end{equation}
where $S(z)$ is the selection function
\begin{equation} \label{selfunc}
S(z) = \int^{\min[M_{\rm max}(z), M_2]}_{\max[M_{\rm min}(z), M_1]}
            \phi_{obs}(M) dM
     \left/ 
       \int^{M_2}_{M_1} \phi(M) dM
     \right. \ ,
\end{equation}
and $w(z)$ is the weight function 
\begin{equation} \label{eqwt}
w(z) = \frac {1} {1 + \bar{f} \bar\rho J_3 S(z)} \ ,
\end{equation}
with 
\begin{equation}
J_3 = 4 \pi \int^\infty_0 r^2 \xi(r) dr
\end{equation}
denoting the second moment of the two-point spatial correlation function
$\xi(r)$. We use $J_3 \approx 10000 \ (h^{-1}$~Mpc)$^3$ as found from a fit 
to the LCRS real space correlation function (\cite{lin96sig}). 
Recall $\phi_{obs}$ is the Schechter function convolved with a
Gaussian magnitude error distribution, and the weights $W_i$ account for
the sampling and completeness effects (2)-(4) of the previous subsection.
Formally we should also incorporate the completeness effects (3) and (4)
into the definition of $S(z)$ above, but since these completeness
corrections are generally not too different from unity, it is simpler
to use the weights $W_i$ instead.
Also, for simplicity, we just use the mean sampling fraction $\bar{f}$
in the denominator of $w(z)$ and ignore the field-to-field variations.

The mean density $\bar\rho$ is calculated iteratively using 
equations~(\ref{eqrho}) 
and (\ref{eqwt}). However, we find that for the redshift and absolute
magnitude limits we are interested in, 1000 km~s$^{-1} < cz <$ 60000 
km~s$^{-1}$
and $-23.0 < M < -17.5$, the minimum-variance estimator gives essentially
the same $\bar\rho$ as when we use inverse selection function 
weighting,
\begin{equation}
w(z) = 1 / S(z) \ .
\end{equation}
We will thus adopt this simpler weighting. Nevertheless, we will
estimate errors $\delta \bar\rho$ in the mean density using the
minimum-variance weighting scheme
\begin{equation} \label{eqrhoerr}
\delta \bar\rho \approx 
   \left[
     \frac{ \bar\rho / \bar{f} }
          { \int^{z_2}_{z_1} S(z) w(z) (\frac{dV}{dz}) dz }
   \right]^{1/2} \sim \bar\rho \left[ \frac{J_3}{V} \right]^{1/2} \ .
\end{equation}
(Here $w(z)$ is that of equation~[\ref{eqwt}]. Note also that the dependence
of $\delta \bar\rho$ on the actual value of $J_3$ used is fairly weak.)
However, this error estimate 
accounts just for that part due to galaxy clustering; we add in quadrature
the uncertainty in $\bar\rho$ arising from varying $M^*$ and $\alpha$ along
their joint 1$\sigma$ error ellipse. 
Finally, we can compute the normalization $\phi^*$ in the Schechter function
(\ref{eqschech}) by
\begin{equation}
\phi^* = \frac{ \bar\rho }
              { \int^{M_2}_{M_1} \phi'(M) dM } \ ,
\end{equation}
where $\phi'$ is the Schechter function with $\phi^*$ set to one. As with
$\bar \rho$, the errors in $\phi^*$ will include contributions both from 
galaxy clustering and from uncertainties in $M^*$ and $\alpha$.

\section{Luminosity Function and Mean Density of LCRS Galaxies} \label{lfrholcrs}

\begin{figure}
\plotfiddle{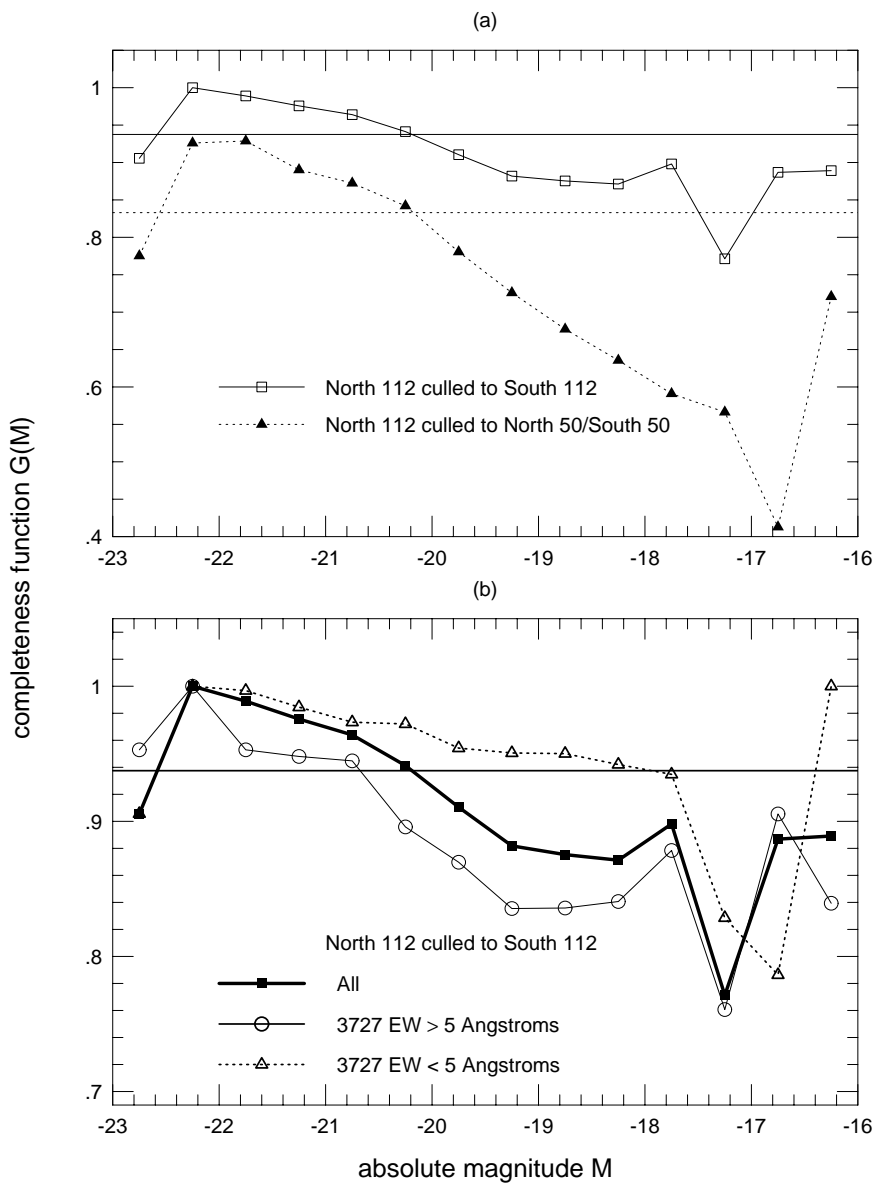}{17cm}{0}{65}{65}{-212}{0}
\caption{(a) Completeness function $G(M)$ vs.\ absolute magnitude $M$. 
          $G(M)$ is 
          calculated from the North 112 sample by culling it to match
          the central surface brightness selection criteria of the 
          North 50/South 50 and South 112 samples. Horizontal lines
          show the expected values of $G(M)$ if the completeness does not
          vary with absolute magnitude. (b) The North 112 sample is again
          culled to the South 112 sample selection criteria, but now $G(M)$ is
          separately plotted for all galaxies and for subsets divided by
          [OII] 3727 equivalent width.}
\label{figG(M)}
\end{figure}

\begin{figure}
\plotfiddle{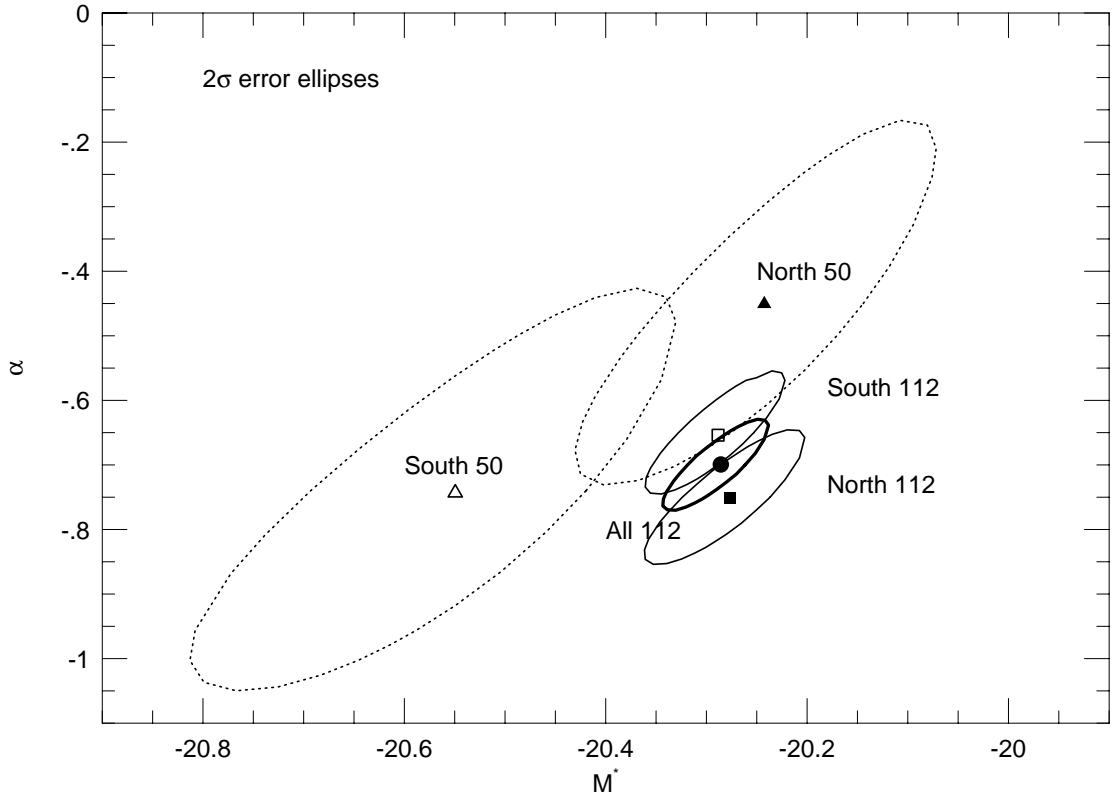}{13cm}{0}{62}{62}{-250}{0}
\caption{2$\sigma$ error ellipses in $M^*$ and $\alpha$ for the STY
          maximum likelihood fits to the four main subsets of the LCRS 
          data and to the combined North+South 112 sample. Details of
          the fits are given in Table~2.}
\label{figco}
\end{figure}

\begin{figure}
\plotfiddle{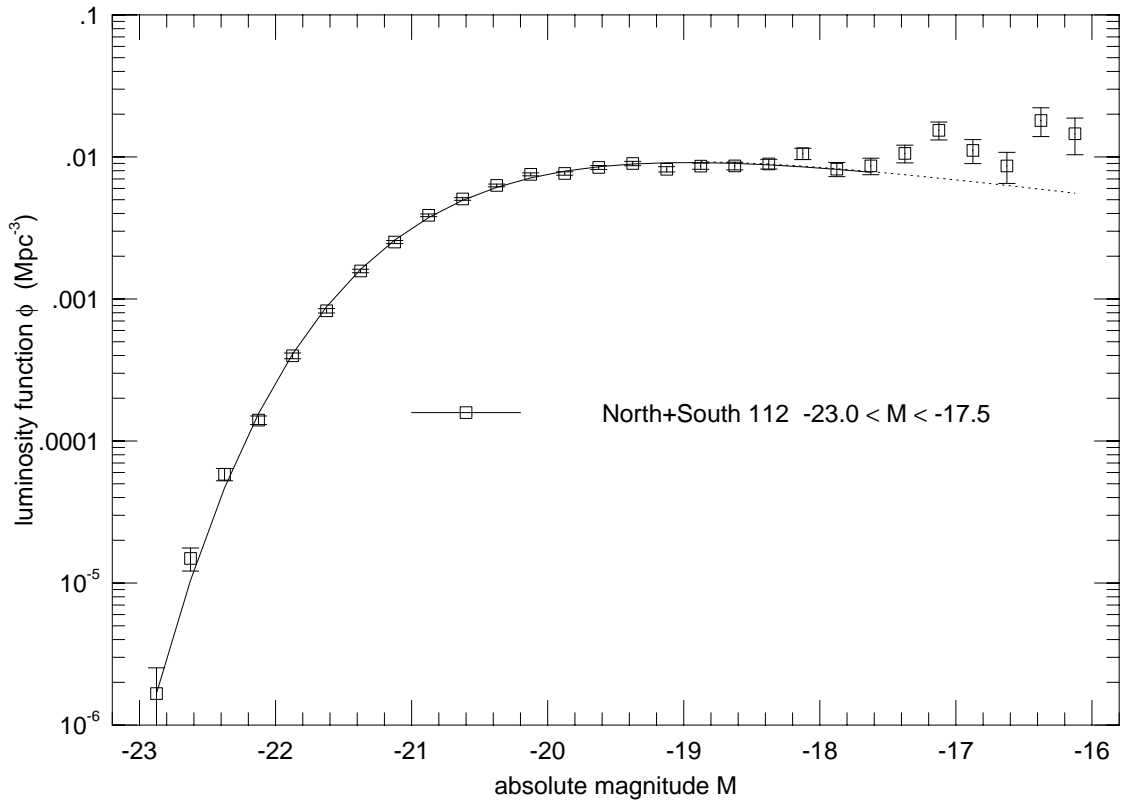}{13cm}{0}{62}{62}{-250}{0}
\caption{Luminosity function for the combined North+South 112 sample.
          The squares show the SWML solution, plotted with 1$\sigma$ errors,
          and the line shows the STY solution.
          Details of the STY fit is given in Table~2. 
          The dotted line shows the extrapolation of the fit to $M > -17.5$.}
\label{figphi}
\end{figure}

\begin{figure}
\plotfiddle{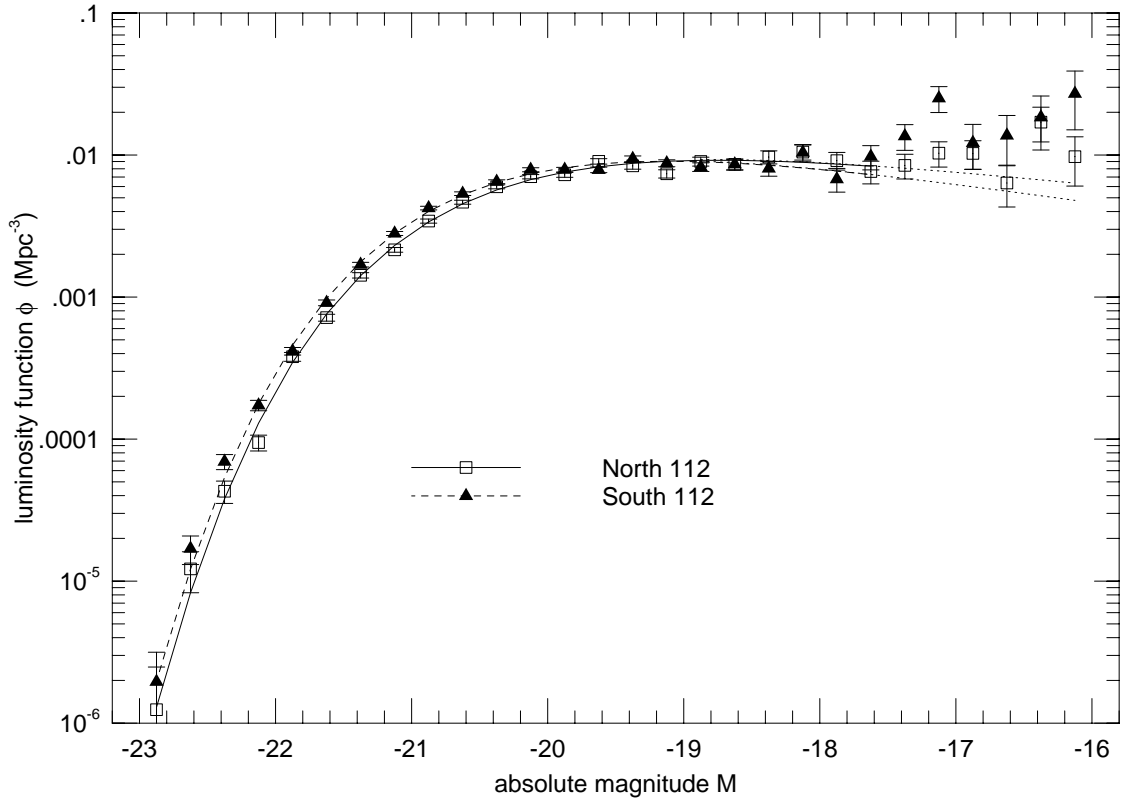}{13cm}{0}{62}{62}{-250}{0}
\caption{Luminosity functions for the North 112 and South 112 samples.
          The points show the SWML solutions, plotted with 1$\sigma$ errors.
          The lines show the STY solutions, with details of the fits given 
          in Table~2.
          Dotted lines show the extrapolation of the fits to $M > -17.5$.}
\label{figphi2}
\end{figure}

Details of the STY fits to the luminosity functions of the N50, S50, N112,
S112, and combined NS112 samples are given in Table~\ref{tabLF}. 
2$\sigma$ error ellipses, calculated using equation~(\ref{ellipse}) with
$\Delta \chi^2 = 6.17$, are plotted in Figure~\ref{figco}. The STY and SWML
solutions for the N112, S112, and NS112 samples are shown in
Figures \ref{figphi} and \ref{figphi2}. We note first that the N50 and S50
results seem discrepant with the NS112 result. We have verified that we can
recover the correct $M^*$ and $\alpha$ parameters, using the methods of
\S~\ref{methods:modifications}, for the N112 and S112 samples culled to 
the 50-fiber limits. It
is thus unclear why the N50 and S50 samples are still discrepant; the 
trends of the discrepancies suggest some residual zero-point offset for the 
S50 sample and some remaining incompleteness in the N50 sample. Since the 
50-fiber data constitute just 20\% of the whole sample, and since they
are subject to larger corrections for central surface brightness selection
(Figure~\ref{figG(M)}a), our final quoted results will use just the 
112-fiber data, even though  
the agreement between the N112 and S112 samples is also not quite ideal.
The N112 and S112 2$\sigma$
error ellipses just touch in Figure~\ref{figco}. The $M^*$ values {\em do}
agree well, but the S112 has an $\alpha$ more positive by 0.1 than that of
N112. This suggests an incompleteness at the faint end of the luminosity 
function, but examination of Figure~\ref{figphi2} shows instead that the
difference arises from an excess of galaxies in the S112 sample at the
bright end $M \lesssim -20$. The explanation for this North/South difference
is also unclear.

We have done extensive work in eliminating residual field-to-field 
photometric zero-point
offsets in our data; details of the procedures used are given in 
Shectman et al. (1996). In brief, relative zero-point offsets between
our drift scans are removed by comparing data in overlapping regions
of the scans and by comparisons to additional CCD calibration frames. 
Moreover, the effects of variable sky brightness and seeing on our 
isophotal magnitudes are dealt with by comparing, on a field-by-field
basis, the offsets between galaxy isophotal and aperture magnitudes (using
a 12$''$-diameter aperture that should be insensitive to seeing
and sky brigtness changes). After the correction procedures, the mean 
photometric zero-points from slice to slice showed no offsets greater
than $\approx 0.015$ magnitude.
It is thus unlikely that zero-point offsets internal to the 
data can account for the differences seen in Figure~\ref{figco}, particularly
the suggestive offset of the S50 sample. Externally, a difference in 
extinction between the North and South galactic caps comes to mind. We
have not made a previous correction for this because 96\% of our fields are at
high galactic latitudes $\vert b \vert \geq 40\arcdeg$. If we apply an
extinction law $\Delta m = -0.07 (\csc \vert b \vert - 1)$ for 
$\vert b \vert < 50\arcdeg$ and $\Delta m = 0$ 
otherwise (\cite{san73}), the changes in
both $M^*$ and $\alpha$ are less than 0.02 for all four data sets. 
An additional extinction of about 0.1 magnitude in the North relative to 
the South {\em could} improve the overlap between the N112 and S112 error
ellipses, but would not help with the 50-fiber results. Another possibility
is that our errors are underestimated. However, we find that the scatter
among subsamples of the four main data sets are in fact consistent with the
estimated error ellipses for the subsamples. On the other hand, the
error estimates do not account for any {\em intrinsic} variations in
the luminosity function parameters with location in the universe.
Thus in particular the N112/S112 discrepancy in $\alpha$ may hint at
such cosmic variance, because of the large separation between the 
samples, but the N50/N112 and S50/S112 differences do not, as these
last two pairs of samples are in close proximity.
Another point to note is
that the image quality for the 112-fiber photometry is appreciably 
improved over that of the early 50-fiber data photometry, due to the improved
CCD's used for the 112-fiber drift scans (\cite{shec95}).
In any event, whether the sample differences are due to some residual
systematic effect(s) or not, we find that
the $M^*$ values for the N50, N112, and S112 samples are in good agreement,
and we will see below in Table~\ref{tabdens} that the galaxy number
and luminosity densities for the four main data sets are also
consistent within their estimated
2$\sigma$ errors. Nevertheless, in light of the difference in the N112
and S112 $\alpha$'s, we will revise our estimate of the error 
in $\alpha$ for the NS112 sample upward from the formal value (see below).

The likelihood-ratio and $\chi^2$ test (\S~\ref{methods:LF})
probabilities $P$ listed in Table~\ref{tabLF} show that the Schechter function
generally gives only formally marginal fits ($P \geq 0.001$) to the luminosity 
function of the LCRS samples.
Nonetheless, a visual inspection of Figures \ref{figphi} and \ref{figphi2}
shows that the Schechter function is still a useful approximation for the
absolute magnitude range $-23.0 \leq M \leq -17.5$. The 
SWML solutions show excesses relative to the Schechter fits at the faint 
end $M > -17.5$.
We adopt as the
LCRS luminosity function the result for the NS112 sample, valid for the
range $-23.0 \leq M \leq -17.5$, with 1$\sigma$ errors:
\begin{eqnarray} \label{eqlfparams}
 M^* & = & -20.29 \pm 0.02 + 5 \log h \nonumber \\
 \alpha & = & -0.70  \pm  0.05 \\ 
 \phi^* & = & 0.019  \pm  0.001 \ h^3 \ {\rm Mpc}^{-3} \ . \nonumber 
\end{eqnarray}
Here we conservatively estimate the uncertainty in $\alpha$ to cover the
range in $\alpha$ derived from the N112 and S112 samples; the formal error
is $\pm 0.03$. Other than this enlargement of the error in $\alpha$, 
we note that the errors quoted above and elsewhere in this paper are formal
values which should account for random uncertainties in the luminosity
function parameters. Systematic uncertainties arising from the
following sources should be kept in mind and were discussed previously:
completeness corrections (\S~\ref{methods:modifications}), extinction and 
photometric calibrations (earlier this section), and cosmology
(\S~\ref{lumdata}).

\begin{figure}
\plotfiddle{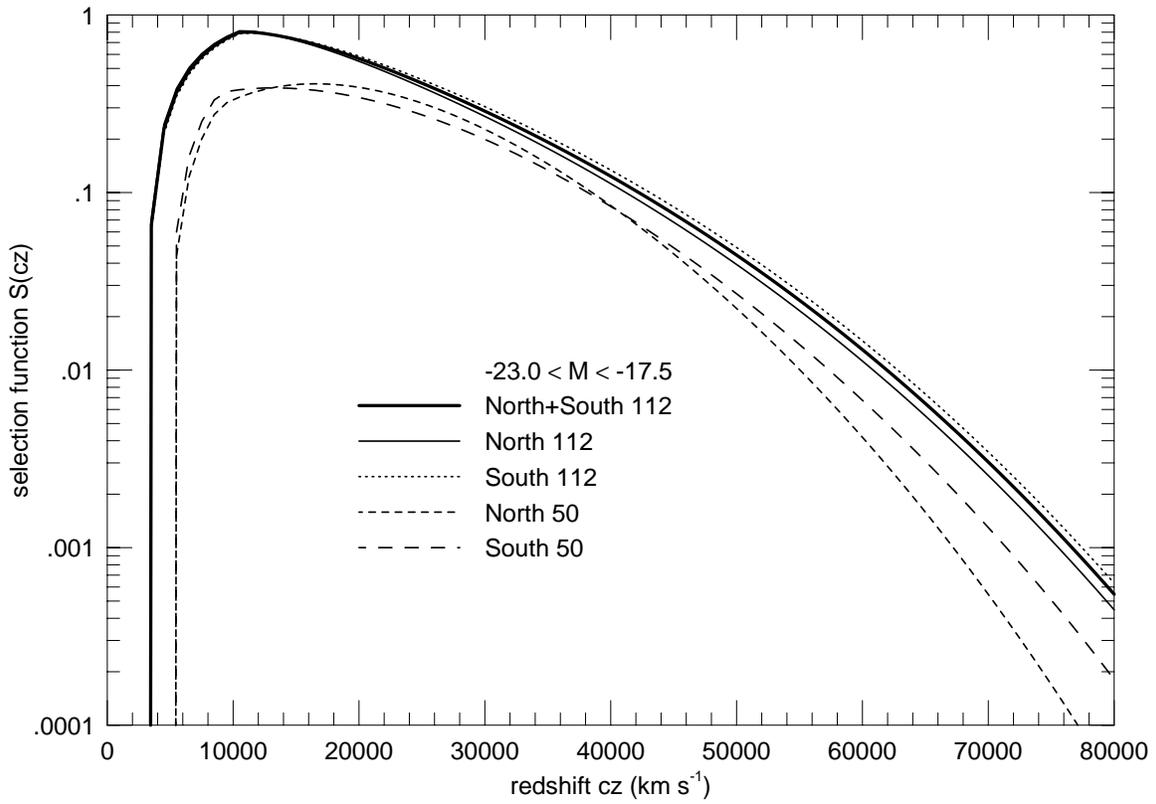}{13cm}{0}{62}{62}{-250}{0}
\caption{The selection function $S(cz)$, computed using the luminosity function
          parameters and nominal apparent magnitude limits of each
          of the North 50, South 50, North 112, South 112, and
          combined North+South 112 samples. Absolute magnitude limits
          $-23.0 \leq M \leq -17.5$ have also been applied.}
\label{figselfunc}
\end{figure}

In our subsequent analyses of the LCRS data (apart from the present paper), 
we will use the luminosity function
values in (\ref{eqlfparams}) for the N50, N112, and S112 data sets, but for 
the S50 data set, we use its own luminosity function parameters as given
in Table~\ref{tabLF}. In Figure~\ref{figselfunc} we plot the selection
function $S(cz)$ computed from equation~(\ref{selfunc}) using the $M^*$ and
$\alpha$ values for each of our four data sets and for the combined NS112 
sample, while restricting the absolute magnitudes to $-23.0 \leq M \leq -17.5$.
Note the typical decline at high $cz$ caused by the survey's faint 
apparent magnitude limits. However, there is also a decline at small $cz$, 
a result of the survey's additional bright apparent magnitude limits.

\begin{figure}
\plotfiddle{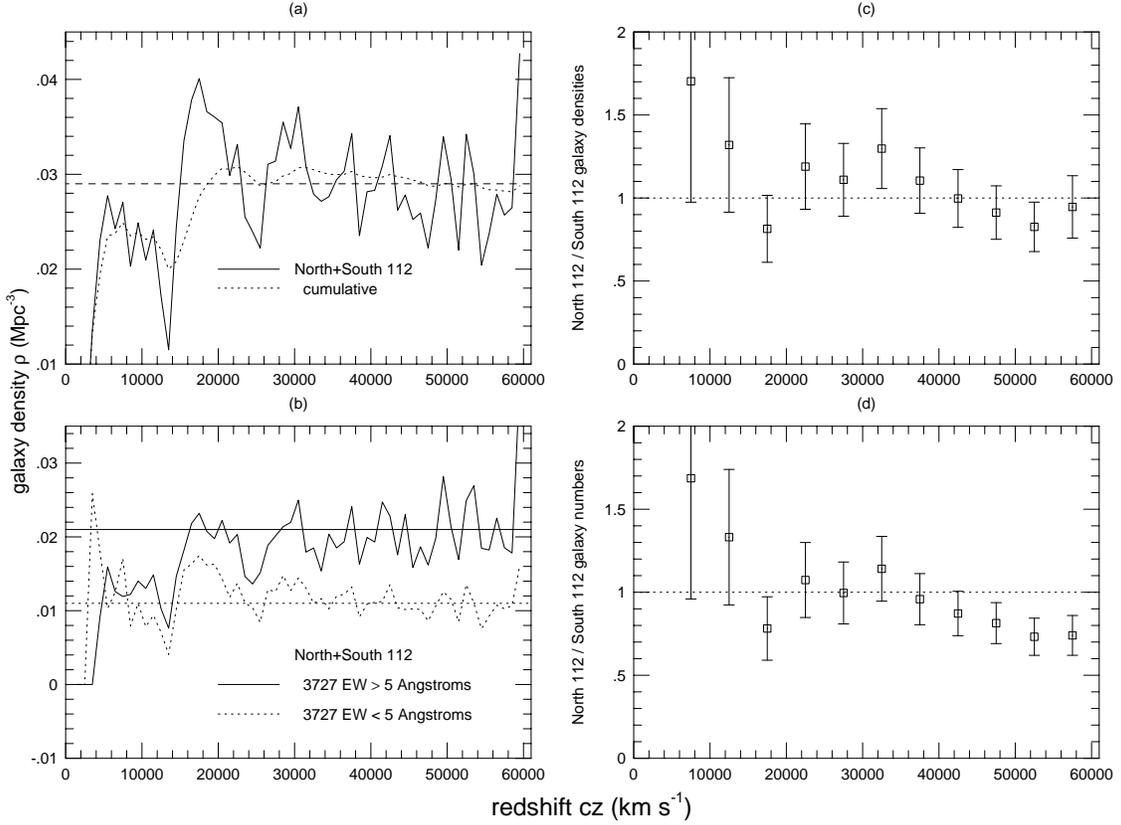}{13cm}{0}{62}{62}{-250}{0}
\caption{The galaxy density $\rho$ as a function of redshift
          $cz$ for: (a) the whole North+South 112 sample; and 
          (b) the North+South 112 sample divided by 3727 emission.
          The horizontal lines show the respective mean densities;
          see Table~3. The ratio of the North 112 over the South 112 
          (c) galaxy number densities and (d) normalized galaxy number 
          histograms are also shown. The errors bars in (c-d) are
          described in the text. In all cases, the absolute magnitude 
          range is $-23.0 \leq M \leq -17.5$.}
\label{figrho}
\end{figure}

The mean densities computed for the LCRS samples are given in 
Table~\ref{tabdens}.
We plot in Figure~\ref{figrho}(a) the galaxy density $\rho$ as a function of
redshift $cz$, for galaxies in the NS112 sample 
within absolute magnitude limits $-23.0 \leq M \leq -17.5$. We also plot
the cumulative density for galaxies lying within redshift $cz$, showing that
the density appears to have converged within the LCRS survey volume. We
stop at $cz = 60000$ km~s$^{-1}$, where the selection function 
$S(cz)$ has dropped
to approximately 0.01 (Figure~\ref{figselfunc}).
We find a mean density $\bar\rho = 0.029 \pm 0.002$ galaxies 
$h^3$~Mpc$^{-3}$. Galaxy
clustering contributes a 4\% fractional uncertainty to $\bar\rho$, 
and uncertainties in the luminosity function parameters contribute
another fractional uncertainty of 4\%. Note from Table~\ref{tabdens} that
despite the difference in $\alpha$ values, the mean densities of the N112 and
S112 samples do in fact agree well. To compare the N112 and S112 
radial density distributions in more detail,
we also plot, as a function of $cz$, the ratio of the N112 to S112 
galaxy number densities (Figure~\ref{figrho}c) and 
galaxy number histograms (Figure~\ref{figrho}d; account has been made of the
differences in volume and completeness).
Note that the ratios differ somewhat in
the two figures, since the density ratios 
include the effect of the differing N112/S112 luminosity functions,
but the histogram ratios do not.
We find from either plot that for radial
shells of thickness 5000 km~s$^{-1}$, fluctuations in galaxy numbers
and densities of about 20\% between N112 and S112 are not uncommon. 
In particular, the decline seen in the N112 sample for $cz \gtrsim
40000$~km~s$^{-1}$ is consistent with the deficit of $M \lesssim M^*$
galaxies observed in the N112 vs. S112 luminosity functions 
(Figure~\ref{figphi2}). On the other hand, we also plot, as error bars 
on the ratios, the expected fluctuations arising from the
combination of galaxy clustering, shot noise, and uncertainties in 
$M^*$ and $\alpha$ (the last only for the density ratios). In
computing the galaxy clustering component of the fluctuations
we have accounted for the geometry
of the shells and used a fit to the LCRS power spectrum
(\cite{lin96ps}; this procedure is preferred over the approximate
error estimate of equation~[\ref{eqrhoerr}], which tends to
give an overestimate given the geometry of the
shells). We see that the observed density fluctuations between N112
and S112 appear to be consistent with the expectations,
despite the 2$\sigma$ discrepancies between the N112/S112
$M^*$-$\alpha$ error ellipses. 
The N112/S112 luminosity functions thus do not differ significantly
in normalization (as described by the galaxy number densities), but
do differ in shape (specifically in $\alpha$).

We also tabulate in Table~\ref{tabdens} the 
galaxy luminosity density $\rho_L$, given by
\begin{equation}
\rho_L = \int_{-23.0}^{-17.5} 10^{-0.4 M^*} \phi(M) dM \ .
\end{equation}
(Letting the integration limits go to infinity only increases $\rho_L$
by 3\%.) 
The NS112 value of 
$(1.4 \pm 0.1) \times 10^8 \ L_{\sun} \ h$~Mpc$^{-3}$, when increased 
by 25\% as a rough isophotal-to-total light correction, 
implies a critical mass-to-light ratio 
$(M/L) \approx 1.6 \times 10^3 \ (M/L)_{\sun}$ to close the
universe, in agreement with previous determinations (e.g., \cite{2:KOSSII};
EEP; \cite{marz94}).
Alternatively, we obtain a cosmological density parameter
$\Omega \approx 0.2 \ [(M/L) \ / \ 300 \ h \ (M/L)_{\sun}]$,
if we set $M / L = 300 \ h \ (M/L)_{\sun}$, a 
typical virial mass-to-light ratio estimated from
clusters and groups (e.g., a recent determination from the 
the CNOC cluster sample gives $M / L = 295 \ h \ (M/L)_{\sun}$ in Gunn $r$;
\cite{car96}).

\begin{figure}
\plotfiddle{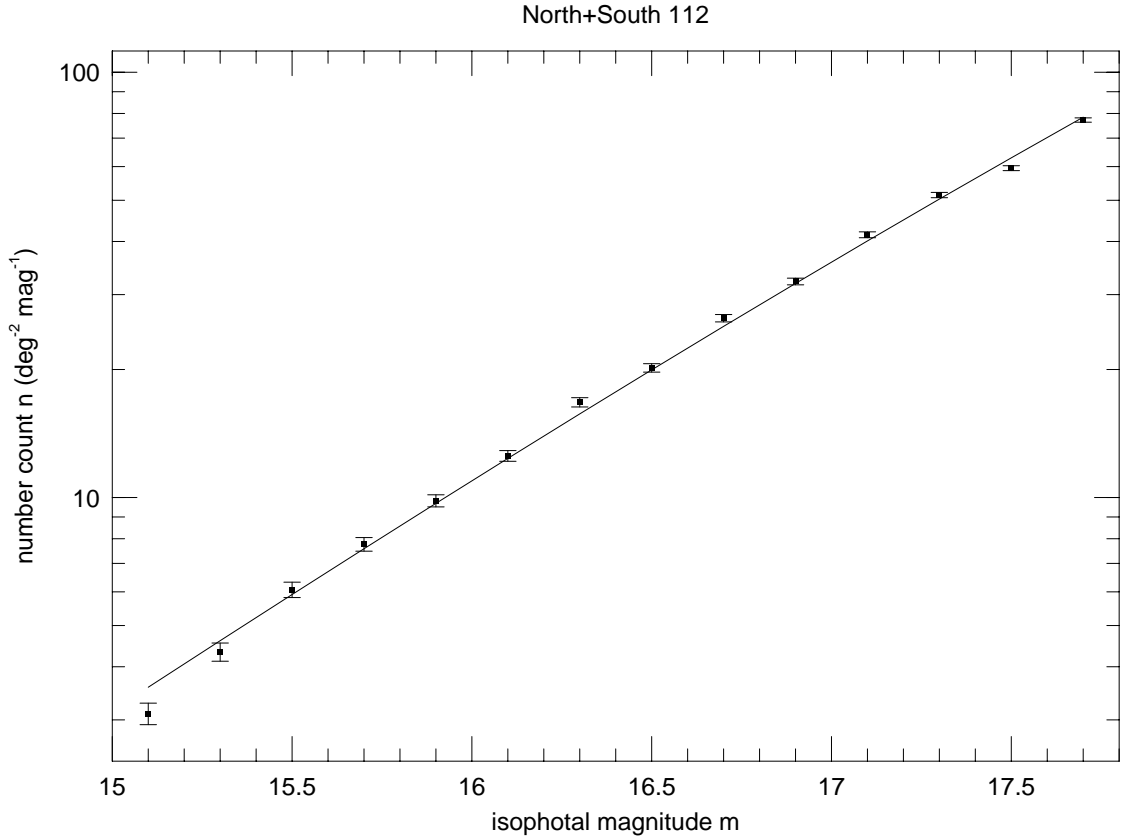}{13cm}{0}{62}{62}{-250}{0}
\caption{The number count $n$ as a function of isophotal magnitude $m$
          for the North+South 112 photometric catalog. Poisson errors
          are shown.
          A 3\% correction has been
          applied to account for stellar contamination. 
          The solid line shows the prediction derived from the NS112
          luminosity function, with normalization fixed by the number
          counts.}
\label{figcounts}
\end{figure}

In addition, we have also derived $\phi^*$ for the NS112 sample by normalizing
to our galaxy number count distribution $dn(m)$. We use
\begin{equation}
\int_m^{m+\Delta m} dn(m) = \int_0^{\infty} dz (dV/dz)
                            \int_M^{M+\Delta M} \phi_{obs}(M') dM' \ ,
\end{equation}
and do a $\chi^2$-fit to find $\phi^*$, assuming Poisson errors on the counts,
and fixing $M^*$ and $\alpha$ at their maximum likelihood values.
For the counts we consider {\em all} objects in our CCD photometric catalog
with isophotal magnitudes $15.0 \leq m \leq 17.8$,
except that we make a 3\% correction
for stellar contamination, as inferred from the fraction of stars in our
spectroscopic sample. We find $\phi^* = 0.021 \ h^3$~Mpc$^{-3}$, but to
properly compare against our redshift survey value, we need to
make a 4\% reduction to account for the low surface brightness galaxies
excluded from
the redshift survey. We get $\phi^* = 0.020 \ h^3$~Mpc$^{-3}$,
in good agreement with the value 
$\phi^* = 0.019 \pm 0.001 \ h^3$~Mpc$^{-3}$ derived from the redshift survey.
We show in Figure~\ref{figcounts} the $dn(m)$ distribution for the NS112
sample and the prediction based on $\phi^* = 0.021 \ h^3$~Mpc$^{-3}$,
$M^* = -20.29$, and $\alpha = -0.70$.

\begin{figure}
\plotfiddle{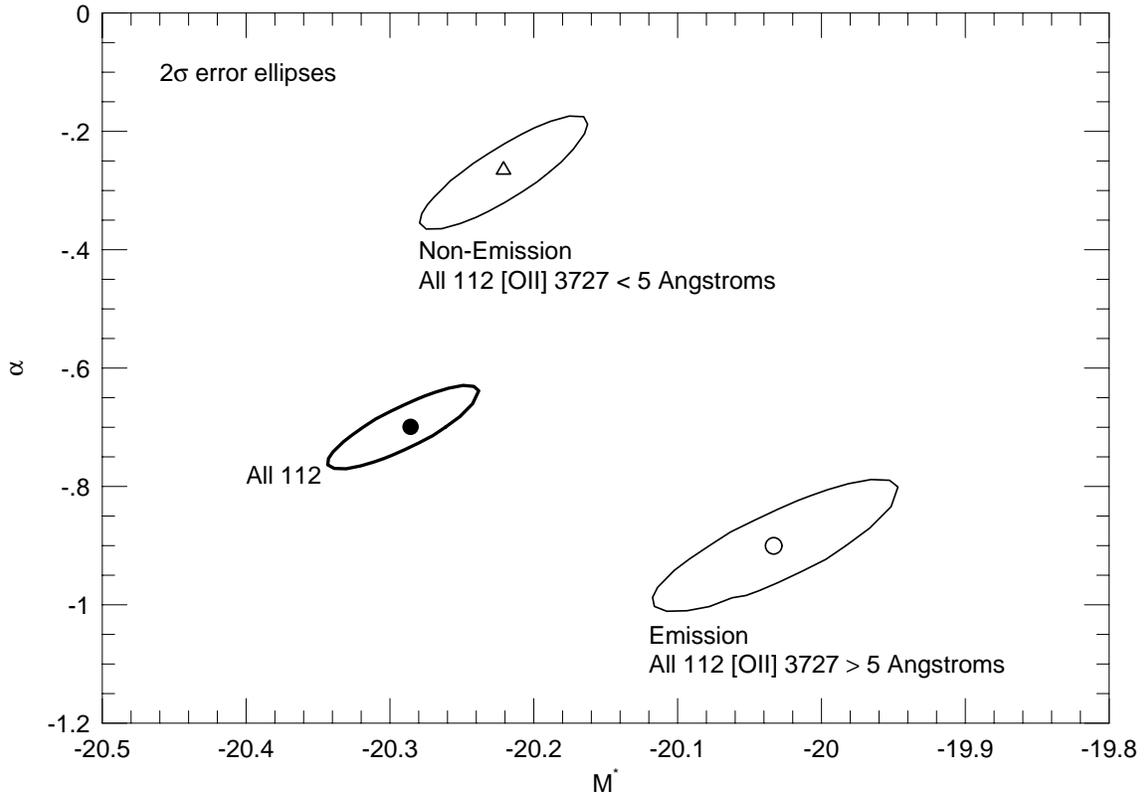}{13cm}{0}{62}{62}{-250}{0}
\caption{2$\sigma$ error ellipses in $M^*$ and $\alpha$ for the STY
          maximum likelihood fits to the 
          combined North+South 112 sample, whole and divided into 2
          subsets by [OII] 3727 equivalent width. Details of
          the fits are given in Table~2.}
\label{figcoem5}
\end{figure}

\begin{figure}
\plotfiddle{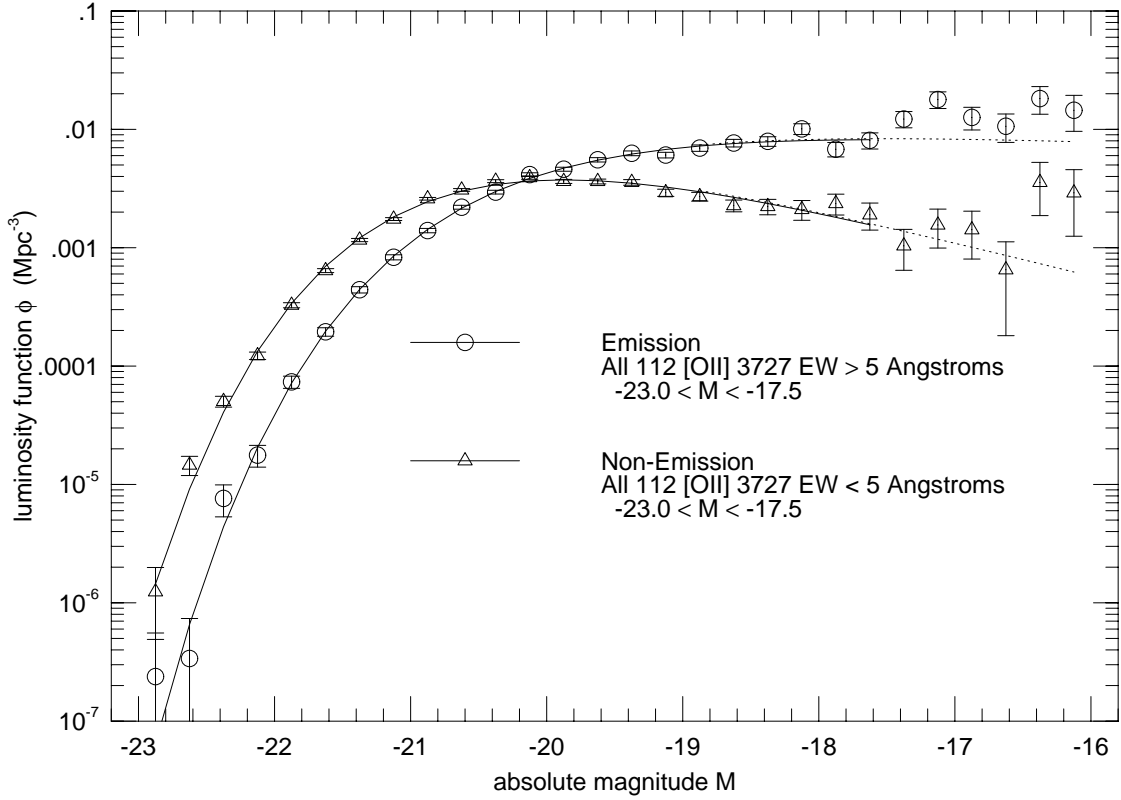}{13cm}{0}{62}{62}{-250}{0}
\caption{Luminosity functions for the combined North+South 112 sample,
          divided by [OII] 3727 equivalent width.
          The points show the SWML solutions, plotted with 1$\sigma$ errors.
          The lines show the STY solutions, with details of the fits given 
          in Table~2.
          Dotted lines show the extrapolation of the fits to $M > -17.5$.}
\label{figphiem5}
\end{figure}

We have also separated the NS112 sample into two groups by [OII] 3727 
emission, defining emission galaxies to be those with 3727 equivalent widths 
$W_{\lambda} \geq 5$~\AA.
The emission galaxies constitute 40\% of the total sample.
The fit details are given in Table~\ref{tabLF}, the error ellipses are
plotted in Figure~\ref{figcoem5}, and the SWML and STY solutions are
shown in Figure~\ref{figphiem5}. Note that for the S112 data, we have used 
separate surface brightness completeness functions $G(M)$, shown in 
Figure~\ref{figG(M)}(b), for the emission and non-emission galaxies.
We see that emission and non-emission
galaxies have very different luminosity functions, with emission galaxies
dominating the faint end and non-emission galaxies prevailing in the 
bright end. In particular, emission galaxies constitute about 15\% of the 
total at $M = -22$, rising to about 80\% at $M = -17$.
The two populations show about equal number densities near the full-sample
$M^*$.
The emission galaxies show a much steeper faint-end slope $\alpha$
($-0.9$ vs.\ $-0.3$) and a slightly fainter $M^*$ ($-20.0$ vs.\ $-20.2$) 
relative
to the non-emission galaxies. Also, note from Figure~\ref{figphiem5} that
the excesses at $M > -17.5$ relative to the Schechter fits are less 
pronounced than they were for the full sample, though for consistency we
use the same fit range $-23.0 \leq M \leq -17.5$. Moreover, the probabilities
in Table~\ref{tabLF} indicate that a Schechter function is a good description
of the emission galaxy luminosity function, though again visual inspection
of Figure~\ref{figphiem5} indicates that the Schechter function provides
reasonable (though not always {\em formally} good) approximations to the 
luminosity functions of both subsamples. In figure~\ref{figrho}(b), we plot
the space density of emission and non-emission galaxies. 
Finally, we note that instead of our formal 1$\sigma$ errors for $\alpha$
listed in Table~\ref{tabLF}, we quote instead an error of $\pm 0.1$ for both
the emission and non-emission galaxies. We do this because of our earlier
N112/S112 difference in $\alpha$, and because of another effect to
be discussed at the end of \S~\ref{counts}.

\section{Comparisons to Other Surveys} \label{comparisons}

\begin{figure}
\plotfiddle{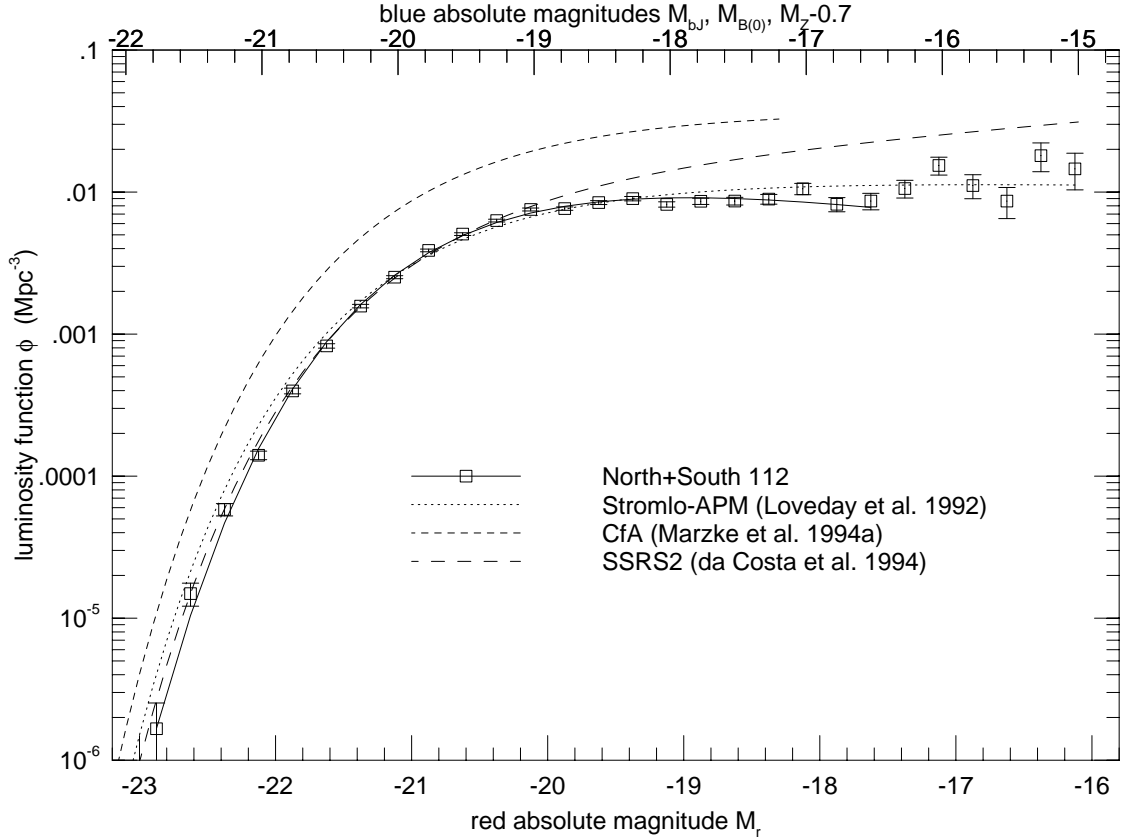}{13cm}{0}{62}{62}{-250}{0}
\caption{The luminosity function for the combined North+South 112 LCRS 
          sample, compared to those of the blue-selected Stromlo-APM, 
          CfA, and SSRS2 surveys. Note that the blue $M_{b_J}$ and
          $M_{B(0)}$ (top) and red $M_r$ (bottom) absolute magnitude
          scales are offset in order to match the average color
          $\langle b_J - R \rangle_0 = 1.1$ of LCRS galaxies. The
          blue Zwicky $M_Z$ magnitude scale has been shifted to 
          match the CfA and Stromlo-APM $M^*$ values. The CfA results
          show an excess relative to the plotted fit at $M_Z \gtrsim
          -16$; see Marzke et al. (1994a).}
\label{figphicomp}
\end{figure}

We next compare our luminosity function to that 
from three other large, local, and optically-selected redshift surveys.
The luminosity functions from the Stromlo-APM (\cite{lov92}),
Center for Astrophysics (CfA; \cite{marz94}), and 
second Southern Sky (SSRS2; \cite{daC94}) redshift surveys are detailed in
Table~\ref{tabLF} and are plotted against
that of the LCRS NS112 sample in Figure~\ref{figphicomp}. Because these
other surveys were selected in the blue, while the LCRS was selected
in the red, for the comparison we need to shift the zero-points of the various
magnitude systems. (Shifting the zero-points also roughly accounts for other
differences in the magnitude systems; e.g., the fact that LCRS
magnitudes are explicitly isophotal whereas APM magnitudes have an
isophotal-to-total magnitude correction applied. 
However, applying simple constant offsets does neglect more complicated
effects, due for example to galaxy colors and morphologies.
Be aware that the specific differences we find below in the luminosity functions
will necessarily be sensitive to the exact shifts in absolute magnitudes we apply.)
For Stromlo-APM we have offset its $M_{b_J}$ magnitude scale by
1.1 magnitude to approximately match 
the mean rest-frame color $\langle b_j-R \rangle_0 = 1.1$
of LCRS galaxies (\cite{tuck94}; \cite{tuck95}). We shift the SSRS2
$M_{B(0)}$ magnitude scale by the same amount, since the SSRS2 and $b_J$ 
zero-points agree within 0.2 magnitude (\cite{daC94}). 
However, we shift the Zwicky
$M_Z$ magnitude scale for the CfA survey by another 0.7 magnitude 
to match the CfA and Stromlo-APM $M^*$ values.
Note first from the figure that the 
Stromlo-APM and LCRS results show consistent normalizations and are fairly 
similar in overall shape, despite differences in their respective Schechter
parameters, particularly in the faint-end slope $\alpha$. 
There are instead differences,
for $M_r \lesssim -21.5$ and $M_r \gtrsim -19$, which are 
responsible for the shallow faint-end slope $\alpha = -0.7$ found in
the LCRS, compared to the more typical flat value $\alpha \approx -1$ seen
in the Stromlo-APM and other previous surveys. 
We emphasize that even
though the LCRS is slightly biased against intrinsically faint galaxies
(because of the survey selection criteria and the correlation between 
luminosity and central surface brightness), the {\it gross} features of our
luminosity function are quite similar to an $\alpha = -1$ Schechter function.
However, such a Schechter function does not provide as good a fit in {\it detail}
to the LCRS data.
As another aid in the comparison, note that for $-1 \leq \alpha \leq -0.7$,
we roughly have $(\Delta \rho_L / \rho_L) \approx 
(\Delta \phi^* / \phi^*) - 0.9 \Delta M^* + \case{1}{3} (\Delta\alpha /
\alpha)$ (good to about 0.1 in $\Delta \rho_L / \rho_L$). Thus the
LCRS and Stromlo-APM surveys
have about the same luminosity density $\rho_L$ despite their different
Schechter parameters (Table~\ref{tabLF}): the 30\% difference in $\phi^*$ is
more than compensated by the 30\% difference in $M^*$ (after noting
the 1.1 magnitude zero-point shift) and one third of the 30\% difference
in $\alpha$.

We note next from Figure~\ref{figphicomp} that the CfA and SSRS2
luminosity functions show some conspicuous differences compared to the LCRS
and Stromlo-APM results. The CfA luminosity function clearly
shows a higher normalization, though its shape is
similar as it has $\alpha = -1$. On the other hand,
the SSRS2 luminosity function is quite similar to those of the LCRS
and Stromlo-APM surveys for $M_{bJ} \lesssim -19$, but for fainter
magnitudes its $\alpha = -1.2$ makes the SSRS2 result overshoot the 
LCRS and Stromlo-APM luminosity functions. Though not shown in the
figure, the CfA luminosity function also has a similar excess at faint
magnitudes $M_Z \gtrsim -16$ relative to an extrapolation of the 
$\alpha = -1$ fit (see Figure 9 of \cite{marz94}). It is unclear why
such a steep faint end
(see below) is not observed in the LCRS or Stromlo-APM surveys.
However, the high normalization of the CfA survey may be explained as
the result of local density inhomogeneities 
(e.g., large coherent structures like the ``Great Wall''; \cite{geller89})
present in the CfA survey volume.
Both the CfA and SSRS2 surveys probe a smaller and more local volume
of space compared to either the LCRS or Stromlo-APM surveys; the mean
survey depths are about 7500 km~s$^{-1}$ for both CfA and SSRS2, 
15000 km~s$^{-1}$ for Stromlo-APM, and 30000 km~s$^{-1}$ for the LCRS.
In fact Marzke et al. (1994a) and da Costa et al. (1994) both point out that
their respective surveys are not yet large enough to be ``fair''
samples of the universe. Both the LCRS and Stromlo-APM surveys should
come closer to representing fair samples, and it is remarkable that
their respective luminosity functions are quite similar in both
normalization and shape.

We do not have morphological classifications for our survey galaxies.
Nonetheless, though there is not a one-to-one correspondence between 3727
emission and morphological type, we do expect our emission galaxies to tend to
be late-type spirals and our non-emission galaxies to be early-type
ellipticals and S0's. We also tabulate in Table~\ref{tabLF} the luminosity
function for galaxies of different morphological types computed for the
Stromlo-APM and CfA (\cite{marz94b}) surveys. The sharp distinction in $\alpha$
between the early- and late-type luminosity functions in the Stromlo-APM survey
($\alpha = +0.2$ and $-0.8$ respectively) qualitatively agrees with the 
emission vs.\ non-emission differences seen in the LCRS. However, 
Zucca et al.\ (1994) and Marzke
et al.\ (1994b) suggest that incomplete classification of early-type galaxies
may be responsible for the difference seen in the Stromlo-APM survey. In
fact, there is no significant difference in the luminosity functions of
E-S0 and spiral (Sa-Sd) luminosity functions in the CfA survey 
($\alpha = -0.8$ to $-0.9$). However, Marzke et al.\ (1994b) do find 
that their faint end is dominated by Magellanic spiral and irregular galaxies,
with a very steep $\alpha = -1.87$. Moreover, da Costa et al. (1994)
attribute their steep $\alpha = -1.2$ to the dominance of blue
galaxies at the faint end. These CfA and SSRS2 results are in
qualitative agreement with the LCRS findings, as we expect 
such late-type blue galaxies to be gas-rich
and to exhibit strong emission lines.

\section{Relation to Galaxy Counts and Galaxy Evolution} \label{counts}

The problem of the excess faint blue galaxy counts over the expectations
of the simplest no-evolution models has continued to be an area of intense 
research (see reviews by \cite{koo92}; \cite{koo95}). The absence in
deep pencil beam redshift surveys (\cite{bro88}; \cite{col90}, 1993; 
\cite{gla95a}) of high redshift $z \gtrsim 1$ galaxies, which are 
predicted by simple luminosity
evolution models for the excess counts, has led to proposals of more exotic
explanations. For example, the excess blue galaxies may be star-bursting
dwarfs which have since faded 
by the present day (e.g., \cite{bro88}; \cite{cow91}; \cite{bab92}), 
or alternatively, extensive galaxy merging 
may have occured to remove the excess galaxy population by $z = 0$
(\cite{bro92}; \cite{roc90}).

In contrast, other workers (\cite{gro95}; \cite{koo93}; \cite{met91}, 1995b)
have argued that the data may still be reasonably fit by 
mild evolution models, given the uncertainties in the faint end slope 
and in the normalization of the local luminosity function; note in particular
the steep $\alpha$ of the Sm-Im sample of Marzke et al.\ (1994b), 
and the difference
in normalization between the Stromlo-APM and CfA surveys. In fact, even
before the CfA results, the counts in the APM galaxy catalog (\cite{mad90a})
at {\em bright} magnitudes 
$b_J = 20.5$ showed a factor of 2 excess relative to no-evolution
predictions based on the counts at $b_J \approx 17$ (\cite{mad90b}). 
Either significant
evolution occurred at {\em low} redshifts $z \approx 0.1$ (\cite{mad90b}),
or there are large-scale structures (voids) and/or magnitude scale errors 
(\cite{met95a}) in the APM survey which are biasing the $b_J \approx 17$
counts systematically low. Choosing to normalize to the galaxy counts 
at $b_J \approx 20$ corresponds to using $\phi^* = 0.03 \ h^3$~Mpc$^{-3}$,
twice the Stromlo-APM value, and using this high normalization reduces
the amount of evolution needed to reconcile with the faint count data.

There is now evidence from deep redshift surveys for evolution in the 
luminosity function since redshifts $z \approx 0.5$ (\cite{eal93};
\cite{lon93}), but one which depends on galaxy type. Blue, star-forming
galaxies appear to exhibit an increased density and/or steepening faint-end
slope with increasing redshift, whereas the red galaxy population shows no
evolution with $z$ (\cite{lil95}; \cite{coll95}). There is also evidence
of an increase in star formation rate with redshift and the association of
merging activity with increased star formation (\cite{coll95};
\cite{col94e}). In particular, the number of star-forming galaxies,
defined as those with [OII] 3727 $W_{\lambda} > 20$~\AA, appears to 
be 5-10 times greater at $z \gtrsim 0.5$ than at $z = 0$, for 
absolute magnitudes below and including $M^*$ (\cite{coll95}).
In addition, recent galaxy number counts from the Hubble Space
Telescope (HST) Medium Deep Survey (\cite{gla95b}; Driver et al.
1995a,b) indicate that
at $B \approx 24$, the counts are dominated by galaxies with irregular and
peculiar morphologies, which show increased numbers relative to their
proportions in local surveys. However, the elliptical and spiral counts are
as expected with no evolution needed, if one adopts the Stromlo-APM 
luminosity function, but with the {\em high} normalization 
$\phi^* = 0.03 \ h^3$~Mpc$^{-3}$.

It is beyond the scope of this paper to undertake detailed 
modelling of the evolution of galaxies, but results from the local
LCRS sample bear on two points which may be connected to understanding the
faint blue galaxies. One is the normalization of the local
luminosity function and the other is the rate of evolution for
galaxies with active star formation.
First, we note that our luminosity function normalization is consistent with
the low value from the Stromlo-APM survey, not with the high value implied 
by the $b_J \approx 20$ APM galaxy counts. Thus results arguing for no or mild
evolution, e.g. the luminosity function models of Gronwall \& Koo (1995)
or the HST elliptical and spiral counts, will need to be modified if our
low normalization is not an underestimate. Because of our large survey depth, up
to $z = 0.2$, large-scale structures should not bias our results low
(cf. Figure~\ref{figrho}). The normalizations for the two completely
independent samples N112 and S112, which probe opposite directions in 
the sky, agree well with each other. Also, the remaining 4\% of low central
surface brightness galaxies that are not part of the redshift survey will
not lead to such a large underestimate. We saw before that our $\phi^*$
normalization is consistent with our total number counts,
{\em including} the low surface brightness galaxies. Any potential factor
of 2 underestimate will have to come from galaxies completely missing from
our original CCD photometric catalog, because they lie below our effective
surface brightness limit. Such a bias against very low surface brightness
galaxies, as has been argued for by, e.g., McGaugh (1994) and Ferguson
\& McGaugh (1995), would be a potential problem for all bright galaxy surveys,
not just the LCRS. We have derived a firm value for the 
density normalization of galaxies with central surface brightnesses 
{\em above} our explicit cut line(s), examples of which are shown in 
Figure~\ref{figsel}.
A final observation here is that our galaxy counts, for our hybrid
Kron-Cousins $R = 15 - 17.8$, 
correspond roughly to $b_J = 16 - 19$. However, whereas the APM counts 
over this range seem
too steep at face value to be consistent with no evolution, even at such 
bright magnitudes (\cite{mad90b}), the Las Campanas counts shown in
Figure~\ref{figcounts} at comparable magnitudes suggest no such evidence
for evolution at low redshifts $z \approx 0.1$.

\begin{figure}
\plotfiddle{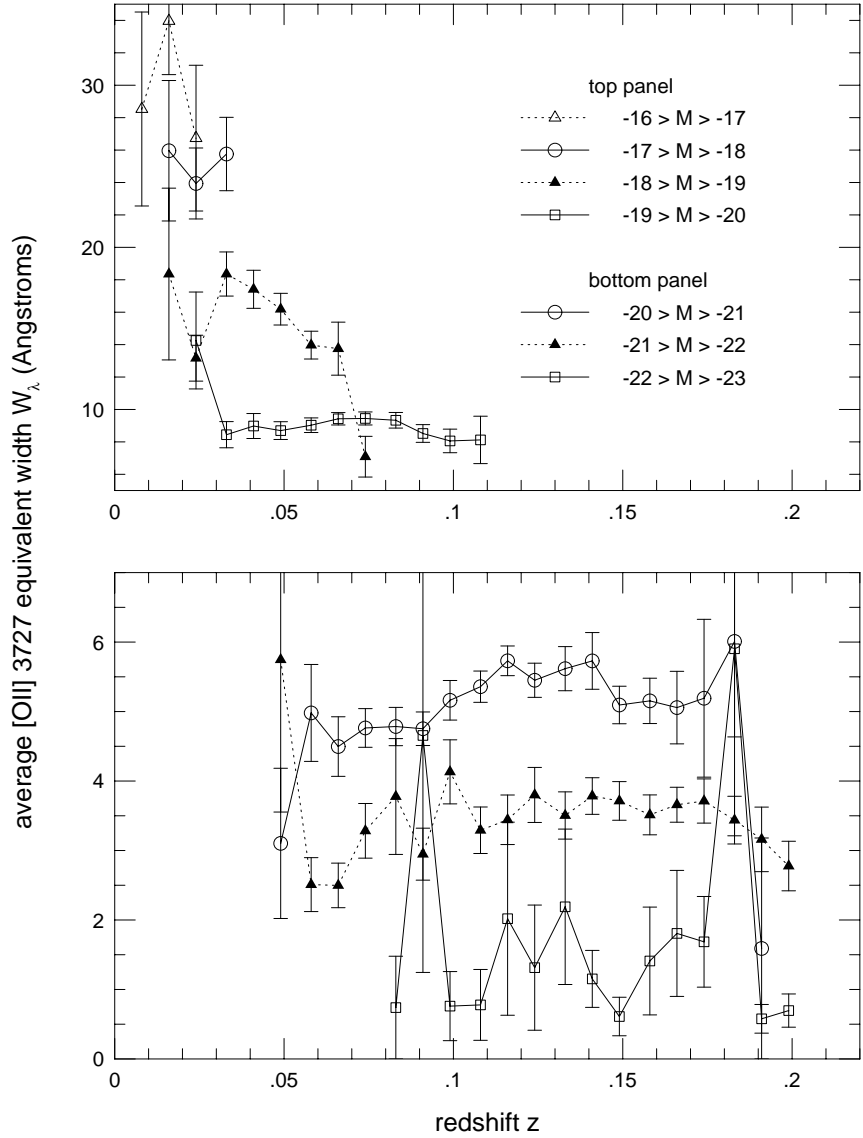}{17cm}{0}{65}{65}{-212}{0}
\caption{The average [OII] 3727 equivalent width as a function of
          redshift, computed for galaxies in the LCRS North+South 112
          sample. The galaxies have been divided into seven bins in
          absolute magnitude, each one magnitude wide, ranging from 
          $M = -16$ to $-23$. The errors shown are standard deviations of
          the mean.}
\label{figeqwave}
\end{figure}

Second, we address the possibility of an increase in the number of
star-forming galxies at redshifts $z < 0.2$ explored by the LCRS, as has
been observed by redshifts $z \approx 0.3$ in deeper surveys 
(\cite{coll95}), by checking whether the average equivalent width of [OII] 3727
varies with redshift. To factor out the dependence of [OII] 3727 $W_{\lambda}$
on galaxy luminosity, we plot in Figure~\ref{figeqwave} the average 
$W_{\lambda}$ as a function of $z$ for seven separate absolute magnitude 
bins, each one magnitude wide, from $M = -16$ to $-23$. We use only our 
NS112 data. Our results indicate that the average equivalent widths generally 
do not change with redshift,
although for galaxies fainter than $M = -19$, the redshift
range is not large. A similar conclusion may also be drawn by plotting the
fraction of galaxies with, say, $W_{\lambda} > 5$~\AA \ or 20~\AA.
In light of the error bars, the most significant trend seen is that for 
the $-20 > M > -21$ bin, specifically a rise of about 1~\AA \
in the average $W_{\lambda}$ at $z \approx 0.1$. This may be due to 
large-scale structure, as the same trend is less apparent in the N112 sample
than in the S112 sample. Alternatively, it may be an aperture effect, in
that our $3.5\arcsec$-diameter fibers cover $5 \ h^{-1}$~kpc on a galaxy
at $z = 0.1$, but only half as much at $z = 0.05$, so that there would be
a bias toward lower equivalent widths at lower redshifts if less emission
arose from the centers of galaxies compared to that from the outer regions.
An argument against the latter explanation is that we do not see any consistent
systematic trend of this kind in the other absolute magnitude bins. In any
event, the evidence for changes in $W_{\lambda}$ is not dramatic in 
any of the luminosity bins.
Since 3727 equivalent width is a surrogate for the star-formation
rate (\cite{ken92}), we conclude that there is no evidence for
significant evolution in the star-forming properties of LCRS galaxies over
the redshift range we sample.
We emphasize that this conclusion is only valid over the {\em respective}
redshift ranges observable for galaxies of different luminosities, and
that galaxies fainter than $M^* + 1 \approx -19$ are not seen above 
$z = 0.1$. For galaxies brighter than $M^*$, though, our no-evolution result 
does extend up to $z = 0.2$. However, note that our null results say nothing
about the above-mentioned evidence 
of increasing numbers of star-forming galaxies at {\em higher} redshifts 
than that constrained by the LCRS, specifically
an increase for sub-$M^*$ galaxies by $z \approx 0.3$ and for $M^*$
galaxies by $z \approx 0.5$ (\cite{coll95}). 

A final note is that the small increase in $W_{\lambda}$ seen for
$M^*$ galaxies does have some impact on the $\alpha$ value derived earlier
for our emission and non-emission galaxy samples. 
In fact, in our earlier derivation we actually
applied redshift-dependent weights, analogous to the completeness
corrections discussed in \S~\ref{methods:modifications}, to account
for any changes in the emission (or non-emission) galaxy
fraction as a function of $z$. The weights are calculated separately
for galaxies of different absolute luminosities.
If we had not applied such
corrections, the maximum-likelihood solution for the emission galaxy 
$\alpha$ would steepen by $-0.1$, while the non-emission
$\alpha$ would flatten by $+0.1$. Leaving out the additional corrections
leads primarily to a systematic increase in the emission galaxy density
with redshift (because of the steeper $\alpha$), 
so that the sum of the cumulative emission and non-emission 
galaxy densities overshoots that for all galaxies by about 25\% at the
$cz = 60000$ km~s$^{-1}$ limit. In light of the lack of evolution seen in the 
Figure~\ref{figeqwave} results, which do {\em not} depend on the luminosity
function, we feel justified in applying the $z$-dependent corrections,
resulting in more reasonable agreement between the total galaxy density and
the sum of the emission and non-emission galaxy densities given in 
Table~\ref{tabdens}.
We increase our estimated errors in the emission and non-emission $\alpha$'s
to $\pm 0.1$ in light of the above effects, as
as we alluded to before at the end of \S~\ref{lfrholcrs}.

\section{Conclusions} \label{conclusions}

We have used inhomogeneity-independent maximum-likelihood methods to
calculate the luminosity function of LCRS galaxies in the (hybrid) Kron-Cousins $R$
absolute magnitude range $-23.0 \leq M - 5 \log h \leq -17.5$, where
we find that it may be reasonably described by a Schechter function with 
$M^* = -20.29 \pm 0.02 + 5 \log h$, $\alpha = -0.70 \pm 0.05$, and 
$\phi^* = 0.019 \pm 0.001 \ h^3$~Mpc$^{-3}$.
There appears to be an 
excess relative to the Schechter function at the faint end $M > -17.5$.
The mean galaxy density for the range $-23.0 \leq M \leq -17.5$ is
$0.029 \pm 0.002$ galaxies $h^3$~Mpc$^{-3}$ out to a redshift of 
$60000$ km~s$^{-1}$,
and the run of cumulative density with redshift suggests that the density
has converged within the LCRS survey volume. 
The measured (isophotal) galaxy luminosity
density is $\rho_L = (1.4~\pm~0.1) \times 10^8 \ L_{\sun} \ h$~Mpc$^{-3}$, resulting
in $\Omega \approx 0.2$, for a typical virial mass-to-light ratio of
$300 \ h \ (M/L)_{\sun}$ and a rough isophotal-to-total light correction.
We have also divided our
sample by the cut [OII] 3727 equivalent width 
$W_{\lambda} \geq 5$~\AA \ , and showed
that the emission and non-emission galaxies so defined 
have significantly different
luminosity functions, with emission galaxies dominating the faint end
($\alpha = -0.9 \pm 0.1$), non-emission galaxies dominating the bright end
($\alpha = -0.3 \pm 0.1$), and about equal numbers near the full-sample $M^*$.

We then compared our results to those from the optically-selected 
Stromlo-APM, CfA, and SSRS2 redshift
surveys. Despite the fact that the Stromlo-APM survey has an
$\alpha \approx -1$, and the LCRS has $\alpha = -0.7$, the shapes of two
luminosity functions are remarkably similar, though they do differ in
specific details. Moreover,
our luminosity function normalization is consistent with that of Stromlo-APM;
the CfA normalization is higher but this may be biased by
density inhomogeneities in its smaller and more local survey volume. 
Finally, our emission galaxy results confirm the picture
seen in these previous surveys that the faint end of the luminosity
function is dominated by late-type, blue galaxies which 
are presumably also gas-rich and strong in [OII] 3727 emission.

Finally, two implications of our results may be made with 
respect to the faint galaxy counts and to galaxy evolution. First,
our normalization of the luminosity function is a factor of two below the
``high'' normalization often adopted from the $b_J \approx 20$ APM galaxy
counts in calculations used to match fainter galaxy count data. If our 
normalization is not an underestimate (e.g., due to
surface brightness effects), then it implies that more evolution
is required in the models, compared to using the high APM count normalization.
Second, for the redshifts sampled in the Las Campanas survey, up to about
$z = 0.2$ for galaxies brighter than $M^*$, our results do not indicate a 
significant change of average [OII] 3727 equivalent width with redshift. 
The star-formation rate, as measured by [OII] 3727, has apparently
not evolved rapidly over the volume sampled by the LCRS, unlike the large
increases seen in the numbers of blue star-forming galaxies in smaller,
deeper surveys at $z \gtrsim 0.3$.

In this paper we have used the Las Campanas Redshift Survey to provide an
accurate determination of the local luminosity function for the largest
galaxy sample thus analyzed. Our red $R$-band results are also complementary
to previous results in blue photometric bands.
In addition, we have begun to explore the luminosity function for
galaxy samples divided by emission properties, and in future analyses
we will examine galaxy samples defined by $b_J - R$ colors (which are 
available for our Southern galaxies from APM observations),
and if possible, also by basic morphological information to be determined
from our drift-scan images. The Las Campanas Redshift Survey should 
thereby provide a useful and large database of local galaxy properties, 
including luminosity, emission, color, and possibly morphological information, 
upon which to anchor models of faint galaxy counts and galaxy evolution 
at higher redshifts.

\acknowledgments

We thank Ron Marzke for useful discussions and the anonymous referee
for helpful suggestions. 
The Las Campanas Redshift Survey has been supported
by NSF grants AST 87-17207, AST 89-21326, and AST 92-20460. HL also
acknowledges support from NASA grant NGT-51093.

Note added in proof: the difference noted here between the luminosity
functions of emission and nonemission galaxies is similar
to that found in preliminary analysis of the ESP redshift
survey galaxies, e.g., Vettolani et al. (1994), IAU Symposium No.
161, 687.

\clearpage

\begin{deluxetable}{lrcccc}

\tablewidth{35pc}
\tablecaption{LCRS Subsamples} 
\tablehead{
\colhead{Sample}   & \colhead{$N$ \tablenotemark{a} } &
\colhead{$\bar{f}$ \tablenotemark{b} } &
\colhead{$m_1$ \tablenotemark{c} }     &
\colhead{$m_2$ \tablenotemark{c} }     &
\colhead{\% LSB \tablenotemark{d} }
 }
\startdata
 North \phn 50-fiber (N50) & 2706 & 0.66 & 16.0 & 17.3 & 22\% \nl
 South \phn 50-fiber (S50) & 2054 & 0.50 & 16.0 & 17.3 & 21\% \nl
 North 112-fiber (N112)   &  8552 & 0.71 & 15.0 & 17.7 & \phn 4\% \nl
 South 112-fiber (S112)   & 10378 & 0.69 & 15.0 & 17.7 & \phn 9\% \nl
\enddata
\tablenotetext{a}{Number of galaxies observed which lie within the
                   survey photometric limits and geometric borders.}
\tablenotetext{b}{Mean sampling fraction for galaxies within photometric
                   limits and geometric borders.}
\tablenotetext{c}{Nominal isophotal magnitude limits: $m_1 \leq m < m_2$.}
\tablenotetext{d}{Percentage of objects excluded for low central surface
                   brightness but otherwise within the isophotal limits.}
\label{tabsamps}
\end{deluxetable}

\clearpage

\begin{deluxetable}{crcccll}
\tablewidth{0pt}
\tablecaption{Luminosity Function Parameters} 
\tablehead{
\colhead{Sample} & \colhead{$N$} & \colhead{$M^* - 5 \log h$ \tablenotemark{a}} &
\colhead{$\alpha$} & \colhead{$\phi^*(h^3$Mpc$^{-3}$) \tablenotemark{a}} &
\colhead{$P_1$ \tablenotemark{b}} & \colhead{$P_2$ \tablenotemark{b}}
 }
\startdata
LCRS \tablenotemark{c} & & & & & & \nl
N50   &   2690 & $-20.24 \pm 0.07$ & $-0.45 \pm 0.11$ 
               & $0.017 \pm 0.003$ & 0.20 & 0.71 \nl
S50   &   2043 & $-20.55 \pm 0.10$ & $-0.74 \pm 0.13$ 
               & $0.016 \pm 0.003$ & 0.03 & 0.04 \nl
N112  &   8399 & $-20.28 \pm 0.03$ & $-0.75 \pm 0.04$ 
               & $0.018 \pm 0.002$ & 0.002 & 0.004 \nl
S112  &  10279 & $-20.29 \pm 0.03 $ & $-0.65 \pm 0.04$ 
               & $0.020 \pm 0.002$ & 0.01 & 0.06 \nl
                       & & & & & & \nl
NS112   &  18678 & $-20.29 \pm 0.02$ & $-0.70 \pm 0.03$ 
                   & $0.019 \pm 0.001$ & 0.001 & 0.009 \nl
3727 $W_{\lambda} \geq 5$~\AA & 7312 & $-20.03 \pm 0.03$ & $-0.90 \pm 0.04$ 
                       & $0.013 \pm 0.001$ & 0.24 & 0.45 \nl
3727 $W_{\lambda} < 5$~\AA &   11366 & $-20.22 \pm 0.02 $ & $-0.27 \pm 0.04 $ 
                       & $0.011 \pm 0.001$ & 0.002 & 0.02 \nl
                       & & & & & & \nl
Stromlo-APM \tablenotemark{d}  & & & & & & \nl
All  & 1658 & $-19.50 \pm 0.13$ & $-0.97 \pm 0.15$
                               & $0.0140 \pm 0.0017$ & 0.65 
                               & \nodata \nl
Early & 311 & $-19.71 \pm 0.25$ & $+0.20 \pm 0.35$ 
                               & \nodata & 0.64 & \nodata \nl
Late  & 999 & $-19.40 \pm 0.16$ & $-0.80 \pm 0.20$ 
                               & \nodata & 0.87 & \nodata \nl
                       & & & & & & \nl
CfA  \tablenotemark{e} & & & & & & \nl
All   & 9063 & $-18.8 \pm 0.3$ & $-1.0 \pm 0.2$ 
                              & $0.04 \pm 0.01$ & 0.25 
                              & \nodata \nl
E-S0  & \nodata & $-18.87$       & $-0.92$
                              & $0.010$ & \nodata & \nodata \nl
Sa-Sd & \nodata & $-18.76$       & $-0.81$ 
                              & $0.015$ & \nodata & \nodata \nl
Sm-Im & \nodata & $-18.79$       & $-1.87$ 
                              & $0.0006 \pm 0.0002$ & 0.46 & \nodata \nl 
                       & & & & & & \nl
SSRS2  \tablenotemark{f} 
   & 2919 & $-19.50$ & $-1.20$ & $0.015 \pm 0.003$ & \nodata
                              & \nodata \nl
\enddata
\tablenotetext{a}{$H_0 = 100 \ h$~km~s$^{-1}$~Mpc$^{-1}$.} 
\tablenotetext{b}{$P_1$ is probability for likelihood ratio test; 
                  $P_2$ is probability for $dN(M)$ $\chi^2$ test. 
                  See \S~3.1.}
\tablenotetext{c}{Magnitudes are in our ``hybrid'' Kron-Cousins $R$ system
                  and absolute
                  magnitudes are restricted to $-23.0 < M - 5 \log h < -17.5$.
                  $\phi^*$ is calculated for redshifts 
                  1000 km~s$^{-1} < cz < 60000$ km~s$^{-1}$. 
                  Errors are formal 1$\sigma$ uncertainties; but also 
                  see discussion in \S\S~4 and 6.}
\tablenotetext{d}{Loveday et al. 1992. Magnitudes in $b_J$ system.}
\tablenotetext{e}{Marzke et al. 1994a,b. Magnitudes in Zwicky system.}
\tablenotetext{f}{da Costa et al. 1994. Magnitudes in $B(0)$ system.}
\label{tabLF}
\end{deluxetable}

\clearpage

\begin{deluxetable}{clc}
\tablewidth{30pc}
\tablecaption{Number and Luminosity Densities}
\tablehead{
\colhead{Sample} &
\colhead{$\bar \rho(h^3$Mpc$^{-3}$) \tablenotemark{a}} &
\colhead{$\rho_L (10^8 L_{\sun} h $Mpc$^{-3}$) \tablenotemark{a,b}} 
 }
\startdata
N50                    & $0.020 \pm 0.004$ & $1.2 \pm 0.1$ \nl
S50                    & $0.026 \pm 0.006$ & $1.5 \pm 0.2$ \nl
N112                   & $0.028 \pm 0.002$ & $1.3 \pm 0.1$ \nl
S112                   & $0.029 \pm 0.002$ & $1.5 \pm 0.1$ \nl
                       &  &  \nl
NS112                  & $0.029 \pm 0.002$ & $1.4 \pm 0.1$ \nl
3727 $W_{\lambda} \geq 5$~\AA        
                       & $0.021 \pm 0.002$ & \phn $0.74 \pm 0.05$ \nl
3727 $W_{\lambda} < 5$~\AA 
                       & $0.011 \pm 0.001$ & \phn $0.78 \pm 0.04$ \nl
\enddata
\tablenotetext{a}{We take $H_0 = 100 \ h$~km~s$^{-1}$~Mpc$^{-1}$. 
                  $\bar \rho$ and $\rho_L$ are calculated for absolute magnitudes
                  $-23.0 < M - 5 \log h < -17.5$ and 
                  redshifts 1000 km~s$^{-1} < cz < 60000$ km~s$^{-1}$.
                  Errors are formal 1$\sigma$ uncertainties.}
\tablenotetext{b}{We assume $R_{\sun} = 4.52$ (M. Pinsonneault,
                  private communication). Also,
                  be aware that our magnitudes are isophotal and are {\em not} total; 
                  we estimate that our isophotal magnitudes miss
                  roughly 25\% of the total light on average.}
\label{tabdens}
\end{deluxetable}

\clearpage

\clearpage

\clearpage

\clearpage

\clearpage

\clearpage

\clearpage

\clearpage

\clearpage

\clearpage

\clearpage

\clearpage

\clearpage

\end{document}